\newif\ifuseTwoColumns
\ifuseTwoColumns
\documentclass[twocolumn]{IEEEtran}
\else
\documentclass[onecolumn,11pt]{IEEEtran}
\usepackage{setspace}
\fi
\usepackage{epsfig}
\usepackage{amsmath}
\usepackage{amssymb}
\usepackage{verbatim}
\usepackage{epic}
\usepackage{graphicx}
\usepackage{signal}

\newcommand{\deft}{{\,\stackrel{\triangle}{=}}\,}

\newcommand{\ste}{\operatorname{s.t.} \,}

\newcommand{\ejo}{{{(e^{j\omega})}}}

\newcommand{\Span}{\operatorname{span} }

\newcommand{\A}{{\mathcal{A}}}

\newcommand{\bphi}{\mbox{\boldmath{$\Phi$}}}
\newcommand{\bal}{\mbox{\boldmath{$\alpha$}}}

\newcommand{\bba}{{{\bf A}}}
\newcommand{\bbz}{{{\bf Z}}}
\newcommand{\bbm}{{{\bf M}}}
\newcommand{\bbw}{{{\bf W}}}

\newcommand{\bbu}{{{\bf U}}}
\newcommand{\bbv}{{{\bf V}}}
\newcommand{\bbb}{{{\bf B}}}
\newcommand{\bbx}{{{\bf X}}}
\newcommand{\bby}{{{\bf Y}}}
\newcommand{\bbi}{{{\bf I}}}

\newcommand{\bbq}{{{\bf Q}}}
\newcommand{\bc}{{{\bf c}}}
\newcommand{\bd}{{{\bf d}}}
\newcommand{\bv}{{{\bf v}}}
\newcommand{\bx}{{{\bf x}}}
\newcommand{\by}{{{\bf y}}}
\newcommand{\bh}{{{\bf h}}}
\newcommand{\bs}{{{\bf s}}}
\newcommand{\bg}{{{\bf g}}}

\newcommand{\bl}{\left(}
\newcommand{\br}{\right)}
\newcommand{\blc}{\left\{}
\newcommand{\brc}{\right\}}

\newcommand{\ZZ}{{\mathbb{Z}}}

\newcommand{\inner}[2]{{\langle#1,#2\rangle}}

\newtheorem{theorem}{Theorem}

\newtheorem{proposition}{Proposition}

\title{Compressed Sensing of Analog Signals in Shift-Invariant Spaces}
\author{Yonina C. Eldar\thanks{Department of Electrical Engineering,
Technion---Israel Institute of Technology, Haifa 32000, Israel.
Phone: +972-4-8293256, fax: +972-4-8295757, E-mail:
yonina@ee.technion.ac.il. This work was supported in part by the
Israel Science Foundation under Grant no. 1081/07 and by the
European Commission in the framework of the FP7 Network of
Excellence in Wireless COMmunications NEWCOM++ (contract no.
216715).}}
\date{\today}

\begin{document}

\maketitle

\begin{abstract}
A traditional assumption underlying most data converters is that
the signal should be sampled at a rate exceeding twice the highest
frequency. This statement is based on a worst-case scenario in
which the signal occupies the entire available bandwidth. In
practice, many signals are sparse so that only part of the
bandwidth is used. In this paper, we develop methods for low-rate
sampling of continuous-time sparse signals in shift-invariant (SI)
spaces, generated by m kernels with period T. We model sparsity by
treating the case in which only k out of the m generators are
active, however, we do not know which k are chosen. We show how to
sample such signals at a rate much lower than m/T, which is the
minimal sampling rate without exploiting sparsity. Our approach
combines ideas from analog sampling in a subspace with a recently
developed block diagram that converts an infinite set of sparse
equations to a finite counterpart. Using these two components we
formulate our problem within the framework of finite compressed
sensing (CS) and then rely on algorithms developed in that
context. The distinguishing feature of our results is that in
contrast to standard CS, which treats finite-length vectors, we
consider sampling of analog signals for which no underlying
finite-dimensional model exists. The proposed framework allows to
extend much of the recent literature on CS to the analog domain.

\end{abstract}

%\newpage
\section{Introduction}

Digital applications have developed rapidly over the last few
decades. Signal processing in the discrete domain inherently
relies on sampling a continuous-time signal to obtain a
discrete-time representation. The traditional assumption
underlying most analog-to-digital converters is that the samples
must be acquired at the Shannon-Nyquist rate, corresponding to
twice the highest frequency \cite{S49,N28}.

Although the bandlimited assumption is often approximately met,
many signals can be more adequately modeled in alternative bases
other than the Fourier basis \cite{D90,U00}, or possess further
structure in the Fourier domain. Research in sampling theory over
the past two decades has substantially enlarged the class of
sampling problems that can be treated efficiently and reliably.
This resulted in many new sampling theories which accommodate more
general signal sets as well as various linear and nonlinear
distortions \cite{AU94,UA94,U00,V01,E02,ED04,DEM08,EM08}.

A signal class that plays an important role in sampling theory are
signals in shift-invariant (SI) spaces. Such functions can be
expressed as linear combinations of shifts of a set of generators
with period $T$ \cite{DDR94,GHM94,CE04,CE05,AG01}. This model
encompasses many signals used in communication and signal
processing. For example, the set of bandlimited functions is SI
with a single generator. Other examples include splines
\cite{U00,S73b} and pulse amplitude modulation in communications.
Using multiple generators, a larger set of signals can be
described such as multiband functions
\cite{LV98,HW99,VB00,ME07,ME08c,ME09}. Sampling theories similar
to the Shannon theorem can be developed for this signal class,
which allows to sample and reconstruct such functions using a
broad variety of filters.

Any signal $x(t)$ in a SI space generated by $m$ functions shifted
with period $T$ can be perfectly recovered from $m$ sampling
sequences, obtained by filtering $x(t)$ with a bank of $m$ filters
and uniformly sampling their outputs at times $nT$. The overall
sampling rate of such a system is $m/T$. In Section~\ref{sec:prob}
we show explicitly how to recover $x(t)$ from these samples by an
appropriate filter bank.  If the signal is generated by $k$ out of
the $m$ generators, then as long as the chosen subset is known, it
suffices to sample at a rate of $k/T$ corresponding to uniform
samples with period $T$ at the output of $k$ filters. However, a
more difficult question is whether the rate can be reduced if we
know that only $k$ of the generators are active, but we do not
know in advance which ones. Since in principle $x(t)$ may be
comprised of any of the generators, it may seem at first that the
rate cannot be lower than $m/T$.

This question is a special case of sampling a signal in a union of
subspaces \cite{LD08,BD09,EM082}. In our problem, $x(t)$ lies in
one of the subspaces expressed by $k$ generators, however we do
not know which subspace is chosen. Necessary and sufficient
conditions where derived in \cite{LD08,BD09} to ensure that a
sampling operator over such a union is invertible. In our setting
this reduces to the requirement that the sampling rate is at least
$2k/T$. However, no concrete sampling methods where given that
ensure efficient and stable recovery, and no recovery algorithm
was provided from a given set of samples. Finite-dimensional
unions where treated in \cite{EM082}, for which stable recovery
methods where developed. Another special case of sampling on a
union of spaces that has been studied extensively is the problem
underlying the field of compressed sensing (CS). In this setting,
the goal is to recover a length $m$ vector $\bx$ from $p<m$ linear
measurements, where $\bx$ is known to be $k$-sparse in some basis
\cite{D06,CRT06}. Many stable and efficient recovery algorithms
have been proposed to recover $\bx$ in this setting
\cite{MPref,BPref2,MEGref1,MElad,CRT06,CT05,EM082}.

 A
fundamental difference between our problem and mainstream CS
papers is that we aim to sample and reconstruct continuous
signals, while CS focuses on recovery of finite vectors. The
methods developed in the context of CS rely on the finite nature
of the problem and cannot be immediately adopted to
infinite-dimensional settings without discretization or
heuristics. Our goal is to directly reduce the analog sampling
rate, without first requiring the Nyquist-rate samples and then
applying finite-dimensional CS techniques.

Several attempts to extend CS ideas to the analog domain were
developed in a set of conferences papers
\cite{Analog2Info1,Analog2Info2}. However, in both papers an
underlying discrete model was assumed which enabled immediate
application of known CS techniques. An alternative analog
framework is the work on finite rate of innovation
\cite{VMB02,DVB07}, in which $x(t)$ is modeled as a finite linear
combination of shifted diracs (some extensions are given to other
generators as well). The algorithms developed in this context
exploit the similarity between the given problem and spectral
estimation, and again rely on finite dimensional methods.

In contrast, the model we treat in this paper is inherently
infinite dimensional as it involves an infinite sequence of
samples from which we would like to recover an analog signal with
infinitely many parameters. In such a setting the measurement
matrix in standard CS is replaced by a more general linear
operator. It is therefore no longer clear how to choose such an
operator to ensure stability. Furthermore, even if a stable
operator can be implemented, it will result in infinitely many
compressed samples. As standard CS algorithms operate on
finite-dimensional optimization problems, they cannot be applied
to infinite dimensional sequences.

In our previous work, we considered a sparse analog sampling
problem in which the signal $x(t)$ has a multiband structure, so
that its Fourier transform consists of at most $N$ bands, each of
width limited by $B$ \cite{ME07,ME08c,ME09}. Explicit sub-Nyquist
sampling and reconstruction schemes were developed in
\cite{ME07,ME08c,ME09} that ensure perfect recovery of  multiband
signals at the minimal possible rate, without requiring knowledge
of the band locations. The proposed algorithms rely on a set of
operations grouped under a block named continuous-to-finite (CTF).
The CTF, which is further developed in \cite{ME08}, essentially
transforms the continuous reconstruction problem into a finite
dimensional equivalent, without discretization or heuristics. The
resulting problem is formulated within the framework of CS, and
thus can be solved efficiently using known tractable algorithms.

The sampling methods used in \cite{ME07,ME09} for blind multiband
sampling are tailored to that specific setting, and are not
applicable to the more general model we consider here. Our goal in
this paper is to capitalize on the key elements from
\cite{ME07,ME08,ME09} that enable CS of multiband signals and
extend them to the more general SI setting by combining results
from standard sampling theory and CS.
 Although the ideas we present are rooted in our previous work,
their application to more general analog CS is not immediately
obvious. To extend our work, it is crucial to setup the more
general problem treated here in a particular way. Therefore, a
large part of the paper is focused on the problem setup, and
reformulation of previously derived results. We then show
explicitly how signals in a SI union created by $m$ generators
with period $T$, can be sampled and stably recovered at a rate
much lower than $m/T$. Specifically, if $k$ out of $m$ generators
are active, then it is sufficient to use $2k \leq p<m$ uniform
sequences at rate $1/T$, where $p$ is determined by the
requirements of standard CS.

The paper is organized as follows. In Section~\ref{sec:prob} we
provide background material on Nyquist-rate sampling in SI spaces.
Although most of these results are known in the literature, we
review them here since our interpretation of the recovery method
is essential in treating the sparse setting. The sparse SI model
is presented in Section~\ref{sec:union}. In this section we also
review the main elements of CS needed for the development of our
algorithm, and elaborate more on the essential difficulty in
extending them to the analog setting. The difference between
sampling in general SI spaces and our previous work \cite{ME07} is
also highlighted. In Section~\ref{sec:acs} we present our strategy
for CS of SI analog signals. Some examples of our framework are
discussed in Section~\ref{sec:examples}.

\section{Background: Sampling in SI Spaces}
\label{sec:prob}

Traditional sampling theory deals with the problem of recovering
an unknown function $x(t) \in L_2$ from its uniform samples at
times $t=nT$ where $T$ is the sampling period. More generally, the
signal may be pre-filtered prior to sampling with a filter
$s^*(-t)$ \cite{U00,V01,EM08,EC04,AG01}, where $(\cdot)^*$ denotes
the complex conjugate, as illustrated in the left-hand side of
Fig.~\ref{fig:sampling}.
 \setlength{\unitlength}{.09in}
\begin{figure}[h]
\begin{center}
\begin{picture}(64,10)(0,0)
%\put(0,0){\framebox(64,10){}}

\put(0,2){ \put(4,4.5){\st{$x(t)$}{\ar{\filt{$s^*(-t)$}{\ar}}}}
\put(21,4.5){
   \put(0,0){\line(3,2){3}}
   \qbezier(0,2)(2,2)(3,-1)
   \put(2.95,-1.3){\vector(0,-1){0.05}}}
\put(24,1.5){\makebox(0,0){$t=nT$}}
\put(25,4.5){\ar{\filt{$\phi_{SA}^{-1}\ejo$}{\ar{\rmult{\ar{\filt{$a(t)$}{\ar
{\et{$x(t)$}}}}}}}}}

\put(27,6){\makebox(0,0){$c[n]$}}
\put(39,6){\makebox(0,0){$d[n]$}}
\put(43,-.5){\makebox(0,0){$\sum_{n \in \ZZ} \delta(t-nT)$}}
\put(42,1){\vector(0,1){2.5}} }

\end{picture}
\end{center}
\caption{Non-ideal sampling and reconstruction.}
\label{fig:sampling}
\end{figure}
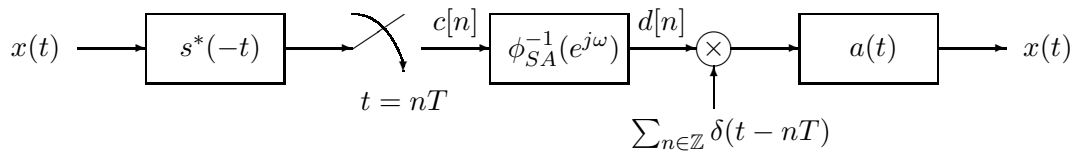
The samples $c[n]$ can be represented as the inner products
$c[n]=\inner{s(t-nT)}{x(t)}$ where
\begin{equation}
\inner{s(t)}{x(t)}=\int_{t = -\infty}^\infty s^*(t)x(t)dt.
\end{equation}
In order to recover $x(t)$ from these samples it is typically
assumed that $x(t)$ lies in an appropriate subspace $\A$ of $L_2$.

A common choice of subspace is a SI subspace generated by a single
generator $a(t)$. Any $x(t) \in \A$ has the form
\begin{equation}
x(t)=\sum_{n \in \ZZ} d[n]a(t-nT),
\end{equation}
for some generator $a(t)$ and sampling period $T$. Note that
$d[n]$ are not necessarily pointwise samples of the signal. If
\begin{equation}
\left| \phi_{SA}\ejo\right |> \alpha>0,\quad \mbox{a.e. } \omega,
\end{equation}
where we defined
\begin{equation}
\label{eq:psa} \phi_{SA}\ejo= \frac{1}{T}\sum_{k \in \ZZ}
S^*\bl\frac{\omega}{T}-\frac{2\pi}{T}k \br
A\bl\frac{\omega}{T}-\frac{2\pi}{T}k \br,
\end{equation}
 then $x(t)$ can be perfectly reconstructed from the
samples $c[n]$ in Fig.~\ref{fig:sampling} \cite{EC04,UA94}. The
function $\phi_{SA}\ejo$ is the discrete-time Fourier transform
(DTFT) of the sampled cross-correlation sequence:
\begin{equation}
r_{sa}[n]=\inner{s(t-nT)}{a(t)}.
\end{equation}
To emphasize the fact that the DTFT is $2\pi$-periodic we use the
notation $\Phi(e^{j\omega})$. Recovery is obtained by filtering
the samples $c[n]$ by a discrete-time filter with frequency
response
\begin{equation}
H(e^{j\omega})=\frac{1}{ \phi_{SA}\ejo},
\end{equation}
followed by modulation by an impulse train with period $T$ and
filtering with an analog filter $a(t)$. The overall sampling and
reconstruction scheme is illustrated in Fig.~\ref{fig:sampling}.
Evidently, SI subspaces allow to retain the basic flavor of the
Shannon sampling theorem in which sampling and recovery are
implemented by filtering operations.

In this paper we consider more general SI spaces, generated by $m$
functions $a_\ell(t) \in L_2,1 \leq \ell \leq m$.
 A
finitely-generated SI subspace in  $L_2$ is defined as
\cite{DDR94,GHM94,CE05}
\begin{equation}
\label{eq:si} \A=\blc x(t)=\sum_{\ell=1}^m \sum_{n \in \ZZ}
d_\ell[n]a_\ell(t-nT): d_\ell[n] \in \ell_2 \brc.
\end{equation}
The functions $a_\ell(t)$ are referred to as the generators of
$\A$. In the Fourier domain, we can represent any $x(t) \in \A$ as
\begin{equation}
\label{eq:xeq} X(\omega)=\sum_{\ell=1}^m D_\ell(e^{j\omega
T})A_\ell(\omega),
\end{equation}
where
\begin{equation}
\label{eq:dft} D_\ell(e^{j\omega })=\sum_{n \in \ZZ}
d_\ell[n]e^{j\omega n},
\end{equation}
is the DTFT of $d_{\ell}[n]$. Throughout the paper, we use
upper-case letters to denote Fourier transforms: $X(\omega)$ is
the continuous-time Fourier transform of the function $x(t)$, and
$C\ejo$ is the DTFT of the sequence $c[n]$.

In order to guarantee a unique stable representation of any signal
in $\A$ by coefficients $d_\ell[n]$, the generators $a_\ell(t)$
are typically chosen to form a Riesz basis for $L_2$. This means
that there exists constants $\alpha>0$ and $\beta<\infty$ such
that
\begin{equation}
\label{eq:riesz} \alpha \|\bd\|^2 \leq \left\| \sum_{\ell=1}^m
\sum_{n \in \ZZ} d_\ell[n]a_\ell(t-nT)\right\|^2\leq \beta
\|\bd\|^2,
\end{equation}
where $\|\bd\|^2=\sum_{\ell=1}^m \sum_{n \in \ZZ} |d_\ell[n]|^2$,
and the norm in the middle term is the standard $L_2$ norm. By
taking Fourier transforms in (\ref{eq:riesz}) it follows that the
generators $a_\ell(t)$ form a Riesz basis if and only if
\cite{GHM94}
\begin{equation}
\label{eq:rc} \alpha \bbi \preceq \bbm_{AA}(e^{j\omega}) \preceq
\beta \bbi,\quad \mbox{a.e. }  \omega,
\end{equation}
where
\begin{equation}
\hspace*{-0.08in} \bbm_{AA}(e^{j\omega})=\left[\begin{array}{ccc}
\phi_{A_1A_1}\ejo & \ldots &
\phi_{A_1A_m}\ejo\\
\vdots & \vdots &  \vdots \\
\phi_{A_mA_1}\ejo & \ldots & \phi_{A_mA_m}\ejo
\end{array}
\right].
\end{equation}
Here $\phi_{A_iA_{\ell}}\ejo$ is defined by (\ref{eq:psa}) with
$A_i\ejo,A_{\ell}\ejo$ replacing $S\ejo,A\ejo$. Throughout the
paper we assume that (\ref{eq:rc}) is satisfied.

Since $x(t)$ lies in a space generated by $m$ functions, it makes
sense to sample it with $m$ filters $s_\ell(t)$, as in the
left-hand side of Fig.~\ref{fig:fbs}.
\setlength{\unitlength}{.105in}
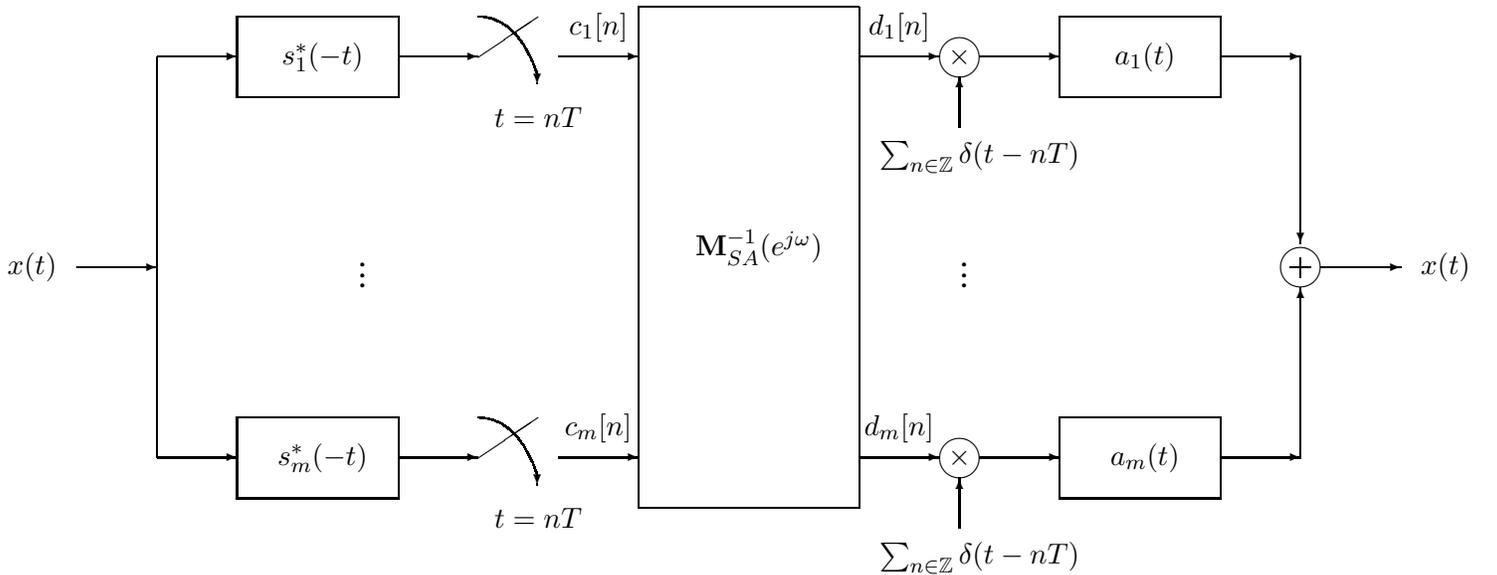
\begin{figure}[h]
\begin{center}
\begin{picture}(77,31)(0,0)
%\put(0,0){\framebox(77,31){}}

\put(-6,0){

 \put(5,0){
\put(5,6.5){{\ar{\filt{$s^*_m(-t)$}{\ar}}}} \put(21,6.5){
   \put(0,0){\line(3,2){3}}
   \qbezier(0,2)(2,2)(3,-1)
   \put(2.95,-1.3){\vector(0,-1){0.05}}}
\put(24,3.5){\makebox(0,0){$t=nT$}} \put(25,6.5){\ar}
\put(40,6.5){\ar{\rmult{\ar{\filt{$a_m(t)$}{\ar}}}}}
\put(27,8){\makebox(0,0){$c_m[n]$}}
\put(46,1.5){\makebox(0,0){$\sum_{n \in \ZZ} \delta(t-nT)$}}
\put(45,3){\vector(0,1){2.5}} \put(42,8){\makebox(0,0){$d_m[n]$}}
}

 \put(20,15){{\Large \vdots}} \put(50,15){{\Large \vdots}}

 \put(5,20){

 \put(5,6.5){{\ar{\filt{$s^*_1(-t)$}{\ar}}}} \put(21,6.5){
   \put(0,0){\line(3,2){3}}
   \qbezier(0,2)(2,2)(3,-1)
   \put(2.95,-1.3){\vector(0,-1){0.05}}}
\put(24,3.5){\makebox(0,0){$t=nT$}} \put(25,6.5){\ar}
\put(40,6.5){\ar{\rmult{\ar{\filt{$a_1(t)$}{\ar}}}}}
\put(27,8){\makebox(0,0){$c_1[n]$}}
\put(46,1.5){\makebox(0,0){$\sum_{n \in \ZZ} \delta(t-nT)$}}
\put(45,3){\vector(0,1){2.5}} \put(42,8){\makebox(0,0){$d_1[n]$}}

}

\put(5,16){\st{$x(t)$}{\ar}} \put(10,6.5){\line(0,1){20}}
\put(34,4){\line(0,1){25}} \put(45,4){\line(0,1){25}}
\put(34,4){\line(1,0){11}} \put(34,29){\line(1,0){11}}
\put(40,17){\makebox(0,0){$\bbm_{SA}^{-1}(e^{j\omega})$}}
\put(66,16){\radder{\ar{\et{$x(t)$}}}}
\put(67,6.5){\vector(0,1){8.5}} \put(67,26.5){\vector(0,-1){9.3}}

}
\end{picture}
\end{center}
\caption{Sampling and reconstruction in shift-invariant spaces.}
\label{fig:fbs}
\end{figure}
 The samples are given by
\begin{equation}
\label{eq:tc} c_\ell[n]=\inner{s_\ell(t-nT)}{x(t)},\quad 1 \leq
\ell \leq m.
\end{equation}
The following proposition provides a simple Fourier-domain
relationship between the samples $c_{\ell}[n]$ of (\ref{eq:tc})
and the expansion coefficients $d_{\ell}[n]$ of (\ref{eq:si}):
\begin{proposition}
\label{prop:sfb} Let $c_\ell[n]=\inner{s_\ell(t-nT)}{x(t)},1 \leq
\ell \leq m$ be a set of $m$ sequences obtained by filtering
$x(t)$ of (\ref{eq:si}) with $m$ filters $s_{\ell}^*(-t)$ and
sampling their outputs at times $nT$, as depicted in the left-hand
side of Fig.~\ref{fig:fbs}. Denote by
$\bc(e^{j\omega}),\bd(e^{j\omega})$ the vectors with $\ell$th
elements $C_\ell(e^{j\omega}),D_\ell(e^{j\omega})$, respectively.
Then
\begin{equation}
\label{eq:sfb}
\bc(e^{j\omega})=\bbm_{SA}(e^{j\omega})\bd(e^{j\omega}),
\end{equation}
where \begin{equation} \label{eq:msa} \hspace*{-0.05in}
\bbm_{SA}(e^{j\omega})= \left[\begin{array}{ccc} \phi_{S_1A_1}\ejo
& \ldots &
\phi_{S_1A_m}\ejo\\
\vdots & \vdots &  \vdots \\
\phi_{S_mA_1}\ejo & \ldots & \phi_{S_mA_m}\ejo
\end{array}
\right],
\end{equation}
and $\phi_{S_iA_{\ell}}\ejo$ are defined by (\ref{eq:psa}).
\end{proposition}
\begin{proof}
The proof follows immediately by taking the Fourier transform of
(\ref{eq:tc}):
\begin{eqnarray}
\label{eq:fc} \lefteqn{\hspace*{-0.25in} C_\ell(e^{j\omega}) =
\frac{1}{T}\sum_{k \in \ZZ} S^*_\ell
\bl\frac{\omega}{T}-\frac{2\pi}{T}k \br
X\bl\frac{\omega}{T}-\frac{2\pi}{T}k \br }\nonumber \\
&\hspace*{-0.58in}  = & \hspace*{-0.35in}\frac{1}{T}\sum_{i=1}^m
D_i(e^{j\omega })\sum_{k \in \ZZ} S^*_\ell
\bl\frac{\omega}{T}-\frac{2\pi}{T}k \br
A_i\bl\frac{\omega}{T}-\frac{2\pi}{T}k \br,
\end{eqnarray}
where we used (\ref{eq:xeq}) and the fact that $D\ejo$ is
$2\pi$-periodic. In vector from, (\ref{eq:fc}) reduces to
(\ref{eq:sfb}).
\end{proof}

Proposition~\ref{prop:sfb} can be used to recover $x(t)$ from the
given samples as long as $\bbm_{SA}(e^{j\omega})$ is invertible
a.e. in $\omega$. Under this condition, the expansion coefficients
can be computed as $\bd\ejo=\bbm^{-1}_{SA}\ejo\bc\ejo$. Given
$d_{\ell}[n]$, $x(t)$ is formed by modulating each coefficient
sequence by a periodic impulse train $\sum_{n \in \ZZ}
\delta(t-nT)$ with period $T$, and filtering with the
corresponding analog filter $a_\ell(t)$. In order to ensure stable
recovery we require $\alpha\bbi \preceq \bbm_{SA}\ejo \preceq
\beta \bbi$ a.e. In particular, we may choose
$s_\ell(t)=a_{\ell}(t)$ due to (\ref{eq:rc}). The resulting
sampling and reconstruction scheme is depicted in
Fig.~\ref{fig:fbs}.

The approach of Fig.~\ref{fig:fbs} results in $m$ sequences of
samples, each at rate $1/T$, leading to an average sampling rate
of $m/T$. Note that from (\ref{eq:si}) it follows that in each
time step $T$, $x(t)$ contains $m$ new parameters, so that the
signal has $m$ degrees of freedom over every interval of length
$T$. Therefore this sampling strategy has the intuitive property
that it requires one sample for each degree of freedom.

\section{Union of Shift-Invariant Subspaces}
\label{sec:union}

Evidently, when subspace information is available, perfect
reconstruction from linear samples is often achievable.
Furthermore, recovery is possible using a simple filter bank. A
more interesting scenario is when $x(t)$ lies in a union of SI
subspaces of the form \cite{EM082}
\begin{equation}
\label{eq:U} x(t) \in \bigcup_{|\ell|=k} \A_\ell,
\end{equation}
where the notation $|\ell|=k$ means a union (or sum) over at most
$k$ elements.
 Here we consider the case in which the union is over $k$
out of $m$ possible subspaces $\{\A_\ell,1 \leq \ell \leq m\}$,
where $\A_\ell$ is generated by $a_\ell(t)$. Thus,
\begin{equation}
\label{eq:model} x(t)=\sum_{|\ell|=k} \sum_{n \in \ZZ}
d_{\ell}[n]a_\ell(t-nT),
\end{equation}
so that only $k$ out of the $m$ sequences $d_{\ell}[n]$ in the sum
(\ref{eq:model}) are not identically zero. Note that
(\ref{eq:model}) no longer defines a subspace. Indeed, while the
union contains signals of the form $\sum_n d_i[n] a_i(t-nT)$ for
two distinct values of $i$, it does not include their linear
combinations.

In principle, if we know which $k$ sequences are non-zero, then
$x(t)$ can be recovered from samples at the output of $k$ filters
using the scheme of Fig.~\ref{fig:fbs}. The resulting average
sampling rate is $k/T$ since we have $k$ sequences, each at rate
$1/T$. Alternatively, even without knowledge of the active
subspaces, we can recover $x(t)$ from samples at the output of $m$
filters resulting in a sampling rate of $m/T$. Although this
strategy does not require knowledge of the active subspaces, the
price is an increase in sampling rate.

In \cite{LD08,BD09}, the authors developed necessary and
sufficient conditions for a sampling operator to be invertible
over a union of subspaces. Specializing the results to our problem
implies that a minimal rate of at least $2k/T$ is needed in order
to ensure that there is a unique SI signal consistent with the
samples. Thus, the fact that we do not know the exact subspace
leads to an increase of at least a factor two in the minimal rate.
However, no concrete methods were provided to reconstruct the
original signal from its samples. Furthermore, although conditions
for invertibility were provided, these do not necessarily imply
that a stable and efficient recovery is possible at the minimal
rate.

Our goal is to develop algorithms for recovering $x(t)$ from a set
of $2k \leq p<m$ sampling sequences, obtained by sampling the
outputs of $p$ filters at rate $1/T$. Before developing our
sampling scheme, we first explain the difficulty in addressing
this problem and its relation to prior work.

%%%%%%%%%%%%%%%%%%%%%%%%%%%%%%%%%%%%%%%%%%%%%%%%%%%
\subsection{Compressed Sensing} \label{sec:cs}
%%%%%%%%%%%%%%%%%%%%%%%%%%%%%%%%%%%%%%%%%%%%%%%%%5

A special case of a union of subspaces that has been treated
extensively is CS of finite vectors \cite{D06,CRT06}. In this
setup, the problem is to recover a finite-dimensional vector $\bx$
of length $m$ from $p<m$ linear measurements $\by$ where
\begin{equation}\label{eq:cs}
\by=\bbm\bx,
\end{equation}
for some matrix $\bbm$ of size $p \times m$. Since the  equations
(\ref{eq:cs}) are underdetermined, more information is needed in
order to recover $\bx$. The prior assumed in the CS literature is
that $\bx=\bphi\bal$ where $\bphi$ is an $m \times m$ invertible
matrix, and $\bal$ is $k$-sparse, so that it has at most $k$
non-zero elements. This prior can be viewed as a union of
subspaces where each subspace is spanned by $k$ columns of
$\bphi$.

A sufficient condition for the uniqueness of a $k$-sparse solution
$\bal$ to the equations $\by=\bbm\bphi\bal$ is that
$\bba=\bbm\bphi$  has a Kruskal-rank of at least $2k$
\cite{MElad,Chen}. The Kruskal-rank is the maximal number $q$ such
that every set of $q$ columns of $\bba$ is linearly independent
\cite{Kruskal}.  This unique $\bal$ can be recovered by solving
the optimization problem \cite{D06}:
\begin{equation}\label{eq:smv0}
\min_{\bal} \|\bal\|_0 \quad \ste \,\by=\bba\bal,
\end{equation}
where the pseudo-norm $\ell_0$ counts the number of non-zero
entries. Therefore, if we are not concerned with stability and
computational complexity, then $2k$ measurements are enough to
recover $\bal$ exactly. Since (\ref{eq:smv0}) is known to be
NP-hard \cite{D06},\cite{CRT06}, several alternative algorithms
have been proposed in the literature that have polynomial
complexity. Two prominent approaches are to replace the $\ell_0$
norm by the convex $\ell_1$ norm, and the orthogonal matching
pursuit algorithm \cite{D06,CRT06}. For a given sparsity level
$k$, these techniques are guaranteed to recover the true sparse
vector $\bal$ as long as certain conditions on $\bba$ are
satisfied, such as the restricted isometry property
\cite{CRT06,C08}. The efficient methods proposed to recover $\bx$
all require a number of measurements $p$ that is larger than $2k$,
however still considerably smaller than $m$. For example, if
$\bba$ is chosen as $p$ random  rows from the Fourier transform
matrix, then the $\ell_1$ program will recover $\bal$ with
overwhelming probability as long as $p \geq c k\log m$ where $c$
is a constant. Other choices for $\bba$ are random matrices
consisting of Gaussian or Bernoulli random variables
\cite{CT05,CR07}. In these cases, on the order of $k\log (m/k)$
measurements are necessary in order to be able to recover $\bal$
efficiently with high probability.

These results have also been generalized to the
multiple-measurement vector (MMV) model in which the problem is to
recover a matrix $\bbx$ from matrix measurements $\bby=\bba\bbx$,
where $\bbx$ has at most $k$ non-zero rows. Here again, if the
Kruskal-rank of $\bba$ is at least $2k$, then there is a unique
$\bbx$ consistent with $\bby$. This unique solution can be
obtained by solving the combinatorial problem
\begin{equation}\label{MMVP0}
\min_{\bbx}|I(\bbx)|,\quad \ste \bby=\bba\bbx,
\end{equation}
where $I(\bbx)$ is the set of indices corresponding to the
non-zero rows of $\bbx$ \cite{Chen}. Various efficient algorithms
that coincide with (\ref{MMVP0}) under certain conditions on
$\bba$ have also been proposed for this problem
\cite{Cotter},\cite{Chen},\cite{ME08}.

%%%%%%%%%%%%%%%%%%%%%%%%%%%%%%%%%%%%%%%%%%%%%%%%%
\subsection{Compressed Sensing of Analog Signals} \label{sec:csa}
%%%%%%%%%%%%%%%%%%%%%%%%%%%%%%%%%%%%%%%%%%%%%%%%%

Our problem is similar in spirit to finite CS: we would like to
sense a sparse signal using fewer measurements than required
without the sparsity assumption. However, the fundamental
difference between the two stems from the fact that our problem is
defined over an infinite-dimensional space of continuous
functions. As we now show, trying to represent it in the same form
as CS by replacing the finite matrices by appropriate operators,
raises several difficulties that precludes direct application of
CS-type results.

To see this, suppose we represent $x(t)$ in terms of a sparse
expansion, by defining an infinite-dimensional operator $\Phi(t)$
corresponding to the concatenation of the functions
$a_{\ell}(t-nT)$, and an infinite sequence $\alpha \in \ell_2$
which consists of the concatenation of the sequences
$d_{\ell}[n]$. We may then write $x(t)=\Phi(t)\alpha$ which
resembles the finite expansion $\bx=\bphi\bal$. Since
$d_{\ell}[n]$ is identically zero for several values of $\ell$,
$\alpha$ will contain many zero elements. Next, we can define a
measurement operator $M(t)$ so that the measurements are given by
$y=A\alpha$ where $A=M(t)\Phi(t)$.

In analogy to the finite setting, the recovery properties of
$\alpha$ should depend on $A$. However, immediate application of
CS ideas to this operator equation is impossible. As we have seen,
the ability to recover $\bal$ in the finite setting depends on its
sparsity. In our case, the sparsity of $\alpha$ is always
infinite. Furthermore, a practical way to ensure stable recovery
with high probability in conventional CS is to draw the elements
of $\bba$ at random, with the number of rows of $\bba$
proportional to the sparsity. In the operator setting, we cannot
clearly define the dimensions of $A$ or draw its elements at
random. Even if we can develop conditions on $A$ such that the
measurement sequence $y=A\alpha$ uniquely determines $\alpha$, it
is still not clear how to recover $\alpha$ from $y$. The immediate
extension of basis pursuit to this context would be:
\begin{equation}
\label{eq:il1} \min_{\alpha} \|\alpha\|_1 \quad \ste \,y=A\alpha.
\end{equation}
Although (\ref{eq:il1}) is a convex problem, it is defined over
infinitely many variables, with infinitely many constraints.
Convex programming techniques such as semi-infinite programming
and generalized semi-infinite programming, allow only for infinite
constraints while the optimization variable must be finite.
Therefore, (\ref{eq:il1}) cannot be solved using standard
optimization tools as in finite-dimensional CS.

This discussion raises three important questions we need to
address in order to adapt CS results to the analog setting:
\begin{enumerate}
\item How do we choose an analog sampling operator?
\item Can we
introduce structure into the sampling operator and still preserve
stability?\item How do we solve the resulting infinite-dimensional
recovery problem?
\end{enumerate}

%%%%%%%%%%%%%%%%%%%%%%%%%%%%%%%%%%%%%%%%%%%%%%%%%
\subsection{Previous Work on Analog Compressed Sensing} \label{sec:csp}
%%%%%%%%%%%%%%%%%%%%%%%%%%%%%%%%%%%%%%%%%%%%%%%%%

In our previous work \cite{ME07,ME08c,ME09}, we treated a special
case of analog CS. Specifically, we considered blind multiband
sampling in which the goal is to sample a bandlimited signal
$x(t)$ whose frequency response consists of at most $N$ bands of
length limited by $B$, with unknown support. In
Section~\ref{sec:examples} we show that this model can be
formulated as a union of SI subspaces (\ref{eq:model}). In order
to sample and recover $x(t)$ at rates much lower than Nyquist, we
proposed two types of sampling methods: In \cite{ME07} we
considered
 multicoset
sampling while in \cite{ME09} we modulated the signal by a
periodic function, prior to standard low-pass sampling. Both
strategies lead to $p$ sequences of low-rate uniform samples
which, in the Fourier domain, can be related to the unknown $x(t)$
via an infinite measurement vector (IMV) model \cite{ME08,ME07}.
This model is an extension of the MMV problem to the case in which
the goal is to recover infinitely many unknown vectors that share
a joint sparsity pattern.  Using the IMV techniques developed in
\cite{ME07,ME08}, $x(t)$ can then be recovered by solving a
finite-dimensional convex optimization problem. We elaborate more
on the IMV model below, as it is a key ingredient in our proposed
sampling strategy.

The sampling method described above is tailored to the multiband
model, and exploits the fact that the spectrum has many intervals
that are identically zero. Applying this approach to the general
SI setting will not lead to perfect recovery. In order to extend
our previous results, we therefore need to reveal the key ideas
that allow CS of analog signals, rather than analyzing a specific
set of sampling equations as in \cite{ME07,ME08c,ME09}. Further
study of blind multiband sampling suggests two key elements
enabling analog CS:
\begin{enumerate}
\item Fourier domain analysis of the sequences of samples. \item
Choosing the sampling functions such that we obtain an IMV model.
\end{enumerate}
Our approach is to capitalize on these two components and extend
them to the model (\ref{eq:model}).

To develop an analog CS system, we design $p<m$ sampling filters
$s_i(t)$ that enable perfect recovery of $x(t)$. In view of our
previous discussion, our task can be rephrased as determining
sampling filters such that in the Fourier domain, the resulting
samples can be described by an IMV system. In the next section we
review the key elements of the IMV problem. We then show how to
appropriately choose the sampling filters for the general model
(\ref{eq:model}).

\section{Infinite Measurement Model} \label{sec:imv}

In the IMV model the goal is to recover a set of unknown vectors
$\bx(\lambda)$ from measurement vectors
\begin{equation}
\label{eq:imv} \by(\lambda)=\bba\bx(\lambda),\quad \lambda \in
\Lambda,
\end{equation}
where $\Lambda$ is a set whose cardinality can be infinite. In
particular, $\Lambda$ may be uncountable, such as the frequencies
$\omega \in [-\pi,\pi)$. The $k$-sparse IMV model assumes that the
vectors $\{\bx(\lambda)\}$, which we denote for brevity by
$\bx(\Lambda)$, share a joint sparsity pattern, that is, the
non-zero elements  are supported on a fixed location set of size
$k$ \cite{ME08}.

As in the finite-case it is easy to see that if $\sigma(\bba) \geq
2k$, where $\sigma(\bba)$ is the Kruskal-rank of $\bba$, then
$\bx(\Lambda)$ is the unique $k$-sparse solution of (\ref{eq:imv})
\cite{ME08}. The major difficulty with the IMV model is how to
recover the solution set $\bx(\Lambda)$ from the infinitely many
equations (\ref{eq:imv}). One suboptimal strategy is to convert
the problem into an MMV by solving (\ref{eq:imv}) only over a
finite set of values $\lambda$. However, clearly this strategy
cannot guarantee perfect recovery. Instead, the approach in
\cite{ME08} is to recover $\bx(\Lambda)$ in two steps. First, we
find the support set $S$ of $\bx(\Lambda)$, and then reconstruct
$\bx(\Lambda)$ from the data $\by(\Lambda)$ and knowledge of $S$.

Once $S$ is found, the second step is straightforward. To see
this, note that using $S$, (\ref{eq:imv}) can be written as
\begin{equation}\label{yAxS}
\by(\lambda) = \bba_S \bx^S(\lambda),\quad\lambda\in\Lambda,
\end{equation}
where $\bba_S$ denotes the matrix containing the columns of $\bba$
whose indices belong to $S$, and $\bx^S(\lambda)$ is the vector
consisting of entries of $\bx(\lambda)$ in locations $S$. Since
$\bx(\Lambda)$ is $k$-sparse, $|S|\leq k$. Therefore, the columns
of $\bba_S$ are linearly independent (because $\sigma(\bba) \geq
2k$), implying that $\bba_S^\dag\bba_S=\bbi$, where $\bba_S^\dag =
\left(\bba_S^H\bba_S\right)^{-1}\bba_S^H$ is the pseudo-inverse of
$\bba_S$ and $(\cdot)^H$ denotes the Hermitian conjugate.
Multiplying (\ref{yAxS}) by $\bba_S^\dag$ on the left gives
\begin{equation}\label{Reconstruct1}
\bx^S(\lambda) = \bba_S^\dag\by(\lambda),\quad\lambda\in\Lambda.
\end{equation}
The components in $\bx(\lambda)$ not supported on $S$ are all
zero. Therefore (\ref{Reconstruct1}) allows for exact recovery of
$\bx(\Lambda)$ once the finite set $S$ is correctly identified.

It remains to determine $S$ efficiently. In \cite{ME08} it was
shown that $S$ can be found exactly by solving a finite MMV. The
steps used to formulate this MMV are grouped under a block
referred to as the continuous-to-finite (CTF) block. The essential
idea is that every finite collection of vectors spanning the
subspace $\Span(\by(\Lambda))$ contains sufficient information to
recover $S$, as incorporated in the following theorem \cite{ME08}:
\begin{theorem}\label{ThKey}
Suppose that $\sigma(\bba) \geq 2k$, and let $\bbv$ be a matrix
with column span equal to $\Span(\by(\Lambda))$. Then, the linear
system
\begin{equation}\label{VAU}
\bbv=\bba\bbu
\end{equation}
has a unique $k$-sparse solution $\bbu$ whose support is equal
$S$.
\end{theorem}
The advantage of Theorem~\ref{ThKey} is that it allows to avoid
the infinite structure of (\ref{eq:imv}) and instead find the
finite set $S$ by solving the single MMV system of (\ref{VAU}).
The additional requirement of Theorem~\ref{ThKey} is to construct
a matrix $\bbv$ having column span equal to $\Span(\by(\Lambda))$.
The following proposition, proven in \cite{ME08}, suggests such a
procedure. To this end, we assume that $\bx(\Lambda)$ is piecewise
continuous in $\lambda$.
\begin{proposition}\label{PropFrame}
If the integral
\begin{equation}\label{MatQ}
\bbq =\int_{\lambda\in\Lambda}\by(\lambda)\by^H(\lambda)d\lambda,
\end{equation}
exists, then every matrix $\bbv$ satisfying $\bbq=\bbv\bbv^H$ has
a column span equal to $\Span(\by(\Lambda))$.
\end{proposition}

\begin{figure} \centering
\includegraphics[scale=1]{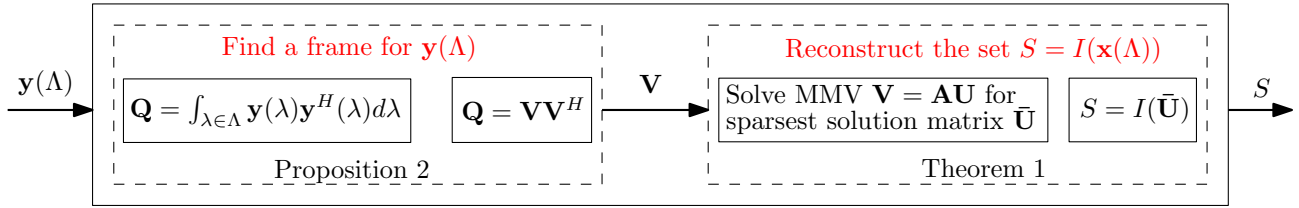}
\caption{The fundamental stages for the recovery of the non-zero
location set $S$ in an IMV model using only one finite-dimensional
problem.}\label{FigIMV}
\end{figure}

 Fig.~\ref{FigIMV}, taken from \cite{ME08},
summarizes the reduction steps that follow from
Theorem~\ref{ThKey} and Proposition~\ref{PropFrame}. Note, that
each block in the figure can be replaced by another set of
operations having an equivalent functionality. In particular, the
computation of the matrix $\bbq$ of Proposition~\ref{PropFrame}
can be avoided if alternative methods are employed for the
construction of a frame $\bbv$ for $\Span(\by(\Lambda))$.
In the figure, $I$ indicates the joint support set of the
corresponding vectors.

%%%%%%%%%%%%%%%%%%%%%%%%%%%%%%%%%%%%%%%%%%%%%%%%%%%
\section{Compressed Sensing of SI Signals}
\label{sec:acs}
%%%%%%%%%%%%%%%%%%%%%%%%%%%%%%%%%%%%%%%%%%%%%%%%%5

We now combine the ideas of Sections~\ref{sec:prob} and
\ref{sec:imv} in order to develop efficient sampling strategies
for a union of subspaces of the form (\ref{eq:model}). Our
approach consists of filtering $x(t)$ with $p<m$ filters $s_i(t)$,
and uniformly sampling their outputs at rate $1/T$. The design of
$s_i(t),1 \leq i \leq p$ relies on two ingredients:
\begin{enumerate}
\item A matrix $\bba$ chosen such that it solves a discrete CS
problem in the dimensions $m$ (vector length) and $k$ (sparsity).
\item A set of functions $h_i(t),1 \leq i \leq m$ which can be
used to sample and reconstruct the entire set of generators
$a_i(t),1 \leq i \leq m$, namely such that
$\bbm_{HA}(e^{j\omega})$ is stably invertible a.e.
\end{enumerate}

The matrix $\bba$ is determined by considering a
finite-dimensional CS problem where we would like to recover a
$k$-sparse vector $\bx$ of length $m$ from $p$ measurements
$\by=\bba\bx$. The value of $p$ can be chosen to guarantee exact
recovery with combinatorial optimization, in which case $p \geq
2k$, or to lead to efficient recovery (possibly only with high
probability) requiring $p>2k$. We show below that the same $\bba$
chosen for this discrete problem can be used for analog CS.  The
functions $h_i(t)$ are chosen so that they can be used to recover
$x(t)$. However, since there are $m$ such functions, this results
in more measurements than actually needed.

We derive the proposed sampling scheme in three steps: First, we
consider the problem of compressively measuring the vector
sequence $\bd[n]$, whose $\ell$th component is given by
$d_\ell[n]$, where only $k$ out of the $m$ sequences $d_\ell[n]$
are non-zero. We show that this can be accomplished by using the
matrix $\bba$ above and IMV recovery theory. In the second step,
we obtain the vector sequence $\bd[n]$ from the given signal
$x(t)$ using an appropriate filter bank of $m$ analog filters, and
sampling their outputs. Finally, we merge the first two steps to
arrive at a bank of $p<m$ analog filters that can compressively
sample $x(t)$ directly. These steps are detailed in the $3$
ensuing subsections.

\subsection{Union of Discrete Sequences}

We begin by treating the problem of sampling and recovering the
sequence $\bd[n]$. This can be accomplished by using the IMV model
introduced in Section~\ref{sec:imv}. Indeed, suppose we measure
 $\bd[n]$ with a size $p \times m$ matrix $\bba$, that allows for CS of $k$-sparse vectors of length $m$.
Then, for each $n$,
\begin{equation}
\label{eq:imvd} \by[n]=\bba\bd[n],\quad n \in \ZZ.
\end{equation}
 The system of (\ref{eq:imvd}) is an IMV model: For
every $n$ the vector $\bd[n]$ is $k$-sparse. Furthermore, the
infinite set of vectors $\{\bd[n],n \in \ZZ\}$ has a joint
sparsity pattern since at most $k$ of the sequences $d_{\ell}[n]$
are non-zero. As we described in Section~\ref{sec:imv}, such a
system of equations can be solved by transforming it into an
equivalent MMV, whose recovery properties are determined by those
of $\bba$. Since $\bba$ was designed such that CS techniques will
work, we are guaranteed that $\bd[n]$ can be perfectly recovered
for each $n$ (or recovered with high probability).
 The reconstruction algorithm is depicted in
Fig.~\ref{FigIMV}. Note that in this case the integral in
computing $\bbq$ becomes a sum over $n$: $\bbq=\sum_{n \in
\ZZ}\by[n]\by^H[n]$ (we assume here that the sum exists).

Instead of solving (\ref{eq:imvd}) we may also consider the
Frequency-domain set of equations:
\begin{equation}
\label{eq:imvf} \by(e^{j\omega})=\bba\bd(e^{j\omega}),\quad 0 \leq
\omega < 2\pi,
\end{equation}
where $\by(e^{j\omega}),\bd(e^{j\omega})$ are the vectors whose
components are the frequency responses
$Y_{\ell}(e^{j\omega}),D_{\ell}(e^{j\omega})$. In principle, we
may apply the CTF block of Fig.~\ref{FigIMV} to either
representations, depending on which choice offers a simpler method
for determining a basis $\bbv$ for the range of
$\{\by(\Lambda)\}$.

When designing the measurements (\ref{eq:imvd}), the only freedom
we have is in choosing $\bba$. To generalize the class of sensing
operators we note that $\bd[n]$ can also be recovered from
\begin{equation}
\label{eq:imvfw}
\by(e^{j\omega})=\bbw(e^{j\omega})\bba\bd(e^{j\omega}),\quad 0
\leq \omega < 2\pi,
\end{equation}
for any invertible $p \times p$ matrix $\bbw(e^{j\omega})$ with
elements $W_{i\ell}(e^{j\omega})$. The measurements  of
(\ref{eq:imvfw}) can be obtained directly in the time domain as
\begin{equation}
\label{eq:timv} y_i[n]=\sum_{\ell=1}^p
w_{i\ell}[n]*\bl\sum_{r=1}^m \bba_{\ell r} d_r[n]\br,\quad 1 \leq
i \leq p,
\end{equation}
where $w_{i\ell}[n]$ is the inverse transform of
$W_{i\ell}(e^{j\omega})$, and $*$ denotes the convolution
operator. To recover  $d_\ell[n]$ from $\by(e^{j\omega})$, we note
that the modified measurements
$\tilde{\by}(e^{j\omega})=\bbw^{-1}\ejo\by(e^{j\omega})$ obey an
IMV model:
\begin{equation}
\label{eq:mimv}
\tilde{\by}(e^{j\omega})=\bba\bd(e^{j\omega}),\quad 0 \leq \omega
< 2\pi.
\end{equation}
Therefore, the CTF block can be applied to $\tilde{\by}\ejo$. As
in (\ref{eq:timv}), we may use the CTF in the time domain by
noting that
\begin{equation}
\label{eq:mimt} \tilde{y}_i[n]=\sum_{\ell=1}^p
b_{i\ell}[n]*y_\ell[n],
\end{equation}
where $b_{i\ell}[n]$ is the inverse DTFT of
$B_{i\ell}(e^{j\omega})$, with
$\bbb(e^{j\omega})=\bbw^{-1}(e^{j\omega})$.

The extra freedom offered by choosing an arbitrary invertible
matrix $\bbw \ejo$ in (\ref{eq:imvfw}) will be useful when we
discuss analog sampling, as different choices lead to different
sampling functions. In Section~\ref{sec:examples} we will see an
example in which a proper selection of $\bbw\ejo$ leads to analog
sampling functions that are easy to implement.

\subsection{Biorthogonal Expansion}

The previous section established that given the ability to sample
the $m$ sequences $d_{\ell}[n]$ we can recover them exactly from
$p<m$ discrete-time sequences obtained via (\ref{eq:imvfw}) or
(\ref{eq:timv}). Reconstruction is performed by applying the CTF
block to the modified measurements either in the frequency domain
(\ref{eq:mimv}) or in the time domain (\ref{eq:mimt}). The
drawback is that we do not have access to $d_{\ell}[n]$ but rather
we are given $x(t)$.

In Fig.~\ref{fig:fbs} and Section~\ref{sec:prob} we have seen that
the sequences $d_{\ell}[n]$ can be obtained by sampling $x(t)$
with a set of functions $h_\ell(t)$ for which $\bbm_{HA}\ejo$ of
(\ref{eq:msa}) is stability invertible, and then filtering the
sampled sequences with a multichannel discrete-time filter
$\bbm_{HA}^{-1} (e^{j\omega})$. Thus, we can first apply this
front-end to $x(t)$, which will produce the sequence of vectors
$\bd[n]$. We can then use the results of the previous subsection
in order to sense these sequences efficiently.  The resulting
measurement sequences $y_\ell[n]$ are depicted in
Fig.~\ref{fig:samp}, where $\bba$ is a $p \times m$ matrix
satisfying the requirements of CS in the appropriate dimensions,
and $\bbw(e^{j\omega})$ is a size $p \times p$ filter bank that is
invertible a.e.
 \setlength{\unitlength}{.09in}
\begin{figure}[h]
\begin{center}
\begin{picture}(74,31)(0,0)
%\put(0,0){\framebox(74,31){}}

 \put(5,0){
\put(5,6.5){{\ar{\filt{$h^*_m(-t)$}{\ar}}}} \put(21,6.5){
   \put(0,0){\line(3,2){3}}
   \qbezier(0,2)(2,2)(3,-1)
   \put(2.95,-1.3){\vector(0,-1){0.05}}}
\put(24,3.5){\makebox(0,0){$t=nT$}} \put(25,6.5){\ar}
\put(57,10){\ar{\et{$y_p[n]$}}}
 }

 \put(20,15){{\Large \vdots}} \put(48,15){{\Large \vdots}}
\put(65,15){{\Large \vdots}}

 \put(5,20){

 \put(5,6.5){{\ar{\filt{$h^*_1(-t)$}{\ar}}}} \put(21,6.5){
   \put(0,0){\line(3,2){3}}
   \qbezier(0,2)(2,2)(3,-1)
   \put(2.95,-1.3){\vector(0,-1){0.05}}}
\put(24,3.5){\makebox(0,0){$t=nT$}} \put(25,6.5){\ar}
\put(57,2){\ar{\et{$y_1[n]$}}}

}

\put(5,16){\st{$x(t)$}{\ar}} \put(10,6.5){\line(0,1){20}}
\put(34,4){\line(0,1){25}} \put(45,4){\line(0,1){25}}
\put(34,4){\line(1,0){11}} \put(34,29){\line(1,0){11}}
\put(40,17){\makebox(0,0){$\bbm_{HA}^{-1}(e^{j\omega})$}}

\put(45,6.5){\lar} \put(48,8){\makebox(0,0){$d_m[n]$}}
\put(45,26.5){\lar} \put(48,28){\makebox(0,0){$d_1[n]$}}

 \put(51,4){\line(0,1){25}}
\put(62,4){\line(0,1){25}} \put(51,4){\line(1,0){11}}
\put(51,29){\line(1,0){11}} \put(57,17){\makebox(0,0){$\bba
\bbw(e^{j\omega})$}}

\end{picture}
\end{center}
\caption{Analog compressed sampling with arbitrary filters
$h_i(t)$.} \label{fig:samp}
\end{figure}
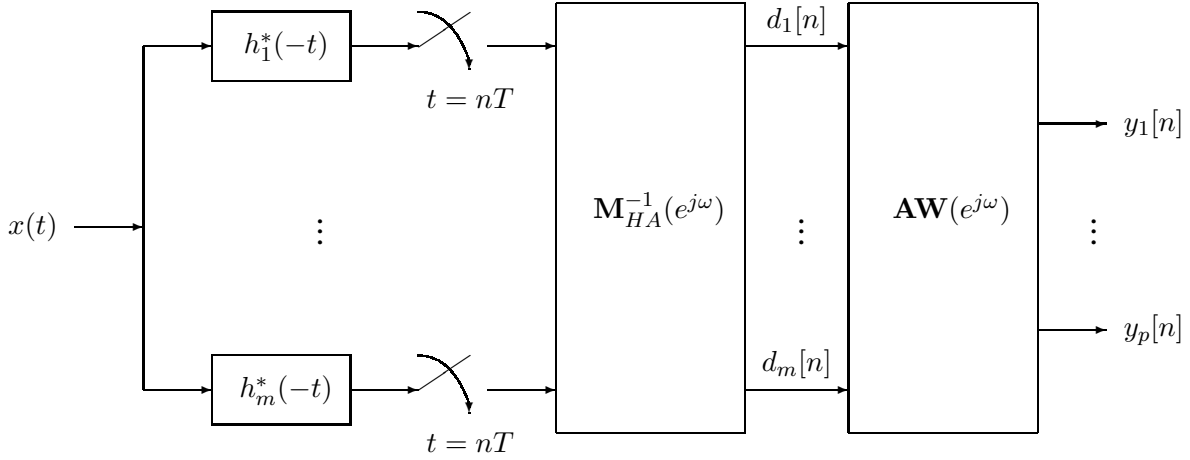

Combining the analog filters $h_{\ell}(t)$ with the discrete-time
multichannel filter $\bbm_{HA}^{-1} (e^{j\omega})$, we can express
$d_{\ell}[n]$ as
\begin{equation}
\label{eq:ci} d_\ell[n]=\inner{v_{\ell}(t-nT)}{x(t)},\quad 1 \leq
\ell \leq m, n \in \ZZ,
\end{equation}
where
\begin{equation}
\label{eq:bios} \bv(\omega)=\bbm_{HA}^{-*} (e^{j\omega
T})\bh(\omega).
\end{equation}
Here $\bv(\omega),\bh(\omega)$ are the vectors with $\ell$th
elements $V_\ell(\omega),H_\ell(\omega)$ and $(\cdot)^{-*}$
denotes the conjugate of the inverse.  The inner products in
(\ref{eq:ci}) can be obtained by filtering $x(t)$ with the bank of
filters $v^*_\ell(-t)$, and uniformly sampling the outputs at
times $nT$.

To see that (\ref{eq:ci}) holds, let $c_{\ell}[n]$ be the samples
resulting from filtering $x(t)$ with the $m$ filters $v_\ell(t)$
and uniformly sampling their outputs at rate $1/T$. From
Proposition~\ref{prop:sfb},
\begin{equation}
\label{eq:cp} \bc(e^{j\omega})=\bbm_{VA}(e^{j\omega}) \bd\ejo.
\end{equation}
Therefore, to establish (\ref{eq:ci}) we need to show that
$\bbm_{VA}(e^{j\omega})=\bbi$. Now, from (\ref{eq:msa}),
\begin{eqnarray}
\label{eq:msan2} \lefteqn{\hspace*{-0.03in}
[\bbm_{VA}(e^{j\omega})]_{i\ell}=\frac{1}{T}\sum_{k \in \ZZ}
V^*_i\bl\frac{\omega}{T}-\frac{2\pi}{T}k \br
A_{\ell}\bl\frac{\omega}{T}-\frac{2\pi}{T}k \br }\nonumber \\
&\hspace*{-0.2in} =& \hspace*{-0.15in}\frac{1}{T}\sum_{r=1}^m
[\bbm^{-1}_{HA}(e^{j\omega})]_{ir} \sum_{k \in \ZZ}
H_{r}^*\bl\frac{\omega}{T}-\frac{2\pi}{T}k \br
A_{\ell}\bl\frac{\omega}{T}-\frac{2\pi}{T}k \br \nonumber \\
& \hspace*{-0.2in} =& \hspace*{-0.15in}
[\bbm_{HA}^{-1}(e^{j\omega})]^i[\bbm_{HA}(e^{j\omega})]_\ell=\bbi_{i\ell},
\end{eqnarray}
where $[\bbq]_{ir}$ is the $ir$th element of the matrix $\bbq$,
and $[\bbq]^i,[\bbq]_i$ are the $i$th row and column respectively
of $\bbq $. Therefore, as required, $\bbm_{VA}(e^{j\omega})=\bbi$.

The functions $\{v_\ell(t-nT)\}$ have the property that they are
biorthogonal to $\{a_\ell(t-nT)\}$, that is
\begin{equation}
\label{eq:bio}
 \inner{v_\ell(t-nT)}{a_i(t-rT)}=\delta_{\ell i}\delta_{nr},
\end{equation}
where $\delta_{\ell i}=1$ if $\ell=i$,  and $0$ otherwise. This
follows from the fact that in the Fourier domain, (\ref{eq:bio})
is equivalent to
\begin{equation}
\label{eq:bio3} \bbm_{VA}\ejo=\bbi.
\end{equation}
 Evidently, we can construct a set of biorthogonal
functions from any set of functions $h_\ell(t)$ for which
$\bbm_{HA}\ejo$ is stably invertible,  via (\ref{eq:bios}). Note
that the biorthogonal vectors in the space $\A$ are unique. This
follows from the fact that if two sets $\{v_i^1(t)\},\{v_i^2(t)\}$
satisfy (\ref{eq:bio}), then
\begin{equation}
\label{eq:biod}
 \inner{g_\ell(t-nT)}{a_i(t-rT)}=0,\quad \mbox{for all }
\ell,i,n,r,
\end{equation}
where $g_i(t)=v_i^1(t)-v_i^2(t)$. Since $\{a_i(t-nT)\}$ span $\A$,
(\ref{eq:biod}) implies that $g_i(t-mT)$ lies in $\A^\perp$ for
any $i,m$. However, if both $v_i^1(t)$ and $v_i^2(t)$ are in $\A$,
then so is $g_i(t-mT)$, from which we conclude that $g_i(t-mT)=0$.
Thus, as long as we start with a set of functions $h_i(t)$ that
span $\A$, the sampling functions $v_i(t)$ resulting from
(\ref{eq:bios}) will be the same. However, their implementation in
hardware is different, since $h_i(t)$ represents an analog filter
while $\bbm_{HA}\ejo$ is a discrete-time filter bank. Therefore,
different choices of $h_i(t)$ lead to distinct analog filters.

\subsection{CS of Analog Signals}

 Although the sampling scheme of Fig.~\ref{fig:samp} results in
compressed measurements $y_\ell[n]$, they are still obtained by an
analog front-end that operates at the high rate $m/T$. However,
our goal is to reduce the rate at the analog front-end. This can
be easily accomplished by moving the discrete filters
$\bbm_{HA}^{-1}(e^{j\omega})$, $\bba\bbw(e^{j\omega})$ back to the
analog domain. In this way, the compressed measurement sequences
$y_\ell[n]$ can be obtained directly from $x(t)$, by filtering
$x(t)$ with $p$ filters $s_{\ell}(t)$ and uniformly sampling their
outputs at times $nT$, leading to a system with sampling rate
$p/T$. An explicit expression for the resulting sampling functions
is given in the following theorem.
\begin{theorem}
\label{thm:fbcs} Let the compressed measurements $y_\ell[n],1 \leq
\ell \leq p$ be the output of the hybrid filter bank in
Fig.~\ref{fig:samp}. Then $\{y_\ell[n]\}$ can be obtained by
filtering $x(t)$ of (\ref{eq:model}) with $p$ filters
$\{s^*_\ell(-t)\}$ and sampling the outputs at rate $1/T$, where
\begin{eqnarray}
\bs(\omega) & = & \bbw^*(e^{j\omega T})\bba^*\bv(\omega) \nonumber
\\
& = & \bbw^*(e^{j\omega T})\bba^*\bbm_{HA}^{-*} (e^{j\omega
T})\bh(\omega).
\end{eqnarray}
Here $\bs(\omega),\bh(\omega)$ are the vectors with $\ell$th
elements $S_\ell(\omega),H_\ell(\omega)$ respectively, and the
components $V_{\ell}(\omega)$ of $\bv(\omega)=\bbm_{HA}^{-*}
(e^{j\omega T})\bh(\omega)$ are Fourier transforms of generators
$v_{\ell}(t)$  such that $\{v_{\ell}(t-nT)\}$ are biorthogonal to
$\{a_\ell(t-nT)\}$. In the time domain,
\begin{equation}
\label{eq:sampt} s_i(t)=\sum_{\ell=1}^m \sum_{r=1}^p \sum_{n \in
\ZZ} w^*_{i r}[-n]\bba^*_{r \ell} v_{\ell}(t-nT),
\end{equation}
where $w_{i r}[n]$ is the inverse transform of $[\bbw\ejo]_{ir}$
and
\begin{equation}
\label{eq:sampv} v_i(t)=\sum_{\ell=1}^m \sum_{n \in \ZZ} \psi^*_{i
\ell}[-n] h_{\ell}(t-nT),
\end{equation}
where $\psi_{i\ell}[n]$ is the inverse transform of
$[\bbm_{HA}^{-1}\ejo]_{i\ell}$.
\end{theorem}
\begin{proof}
Suppose that $x(t)$ is filtered by the $p$ filters $s_i(t)$ and
then uniformly sampled at $nT$. From Proposition~\ref{prop:sfb},
the samples can be expressed in the Fourier domain as
\begin{equation}
\bc(e^{j\omega})=\bbm_{SA}(e^{j\omega}) \bd\ejo.
\end{equation}
In order to prove the theorem we need to show that
$\bbm_{SA}\ejo=\bbw\ejo\bba$.

Let
\begin{equation}
\label{eq:bp} \bbb\ejo=\bbw^*(e^{j\omega})\bba^*,
\end{equation}
so that $\bs(\omega)=\bbb(e^{j\omega T})\bv(\omega)$. Then,
\begin{eqnarray}
\label{eq:msail}
 \lefteqn{[\bbm_{SA}(e^{j\omega})]_{i\ell}=\frac{1}{T}\sum_{k \in \ZZ}
S^*_i\bl\frac{\omega}{T}-\frac{2\pi}{T}k \br
A_{\ell}\bl\frac{\omega}{T}-\frac{2\pi}{T}k \br }\nonumber \\
&=& \frac{1}{T}\sum_{r=1}^m B_{ir}^*(e^{j\omega}) \sum_{k \in \ZZ}
V_{r}^*\bl\frac{\omega}{T}-\frac{2\pi}{T}k \br
A_{\ell}\bl\frac{\omega}{T}-\frac{2\pi}{T}k \br \nonumber \\
&= & [\bbb^*(e^{j\omega})]^i[\bbm_{VA}(e^{j\omega})]_\ell,
\end{eqnarray}
where $[\bbq]^i,[\bbq]_i$ are the $i$th row and column
respectively of the matrix $\bbq$. The first equality follows from
the fact that $\bbb\ejo$ is $2\pi$ periodic. From
(\ref{eq:msail}),
\begin{equation}
\label{eq:msa2t}
\bbm_{SA}(e^{j\omega})=\bbb^*(e^{j\omega})\bbm_{VA}(e^{j\omega})=
\bbw(e^{j\omega})\bba,
\end{equation}
where we used the fact that $\bbm_{VA}(e^{j\omega})=\bbi$ due to
the biorthogonality property.

Finally, if $\bs(\omega)=\bbb(e^{j\omega T}) \bv(\omega)$, then
\begin{equation}
s_i(t)=\sum_{\ell=1}^m \sum_{n \in \ZZ}
b_{i\ell}[n]v_{\ell}(t-nT),
\end{equation}
where $b_{i \ell}[n]$ is the inverse DTFT of
$[B_{i\ell}\ejo]_{i\ell}$. Using (\ref{eq:bp}) together with the
fact that the inverse transform of $Q^*_{i\ell}\ejo$ is
$q^*_{i\ell}[-n]$, results in (\ref{eq:sampt}). The relation
(\ref{eq:sampv}) follows from the same considerations.
\end{proof}

Theorem~\ref{thm:fbcs} is the main result which allows for
compressive sampling of analog signals. Specifically, starting
from any matrix $\bba$ that satisfies the CS requirements of
finite vectors, and a set of sampling functions $h_i(t)$ for which
$\bbm_{HA}\ejo$ is invertible, we can create a multitude of
sampling functions $s_i(t)$ to compressively sample the underlying
analog signal $x(t)$. The sensing is performed by filtering $x(t)$
with the $p<m$ corresponding filters, and sampling their outputs
at rate $1/T$. Reconstruction from the compressed measurements
$y_i[n],1 \leq i \leq p$ is obtained by applying the CTF block of
Fig.~\ref{FigIMV} in order to recover the sequences $d_i[n]$. The
original signal $x(t)$ is then constructed by modulating
appropriate impulse trains and filtering with $a_i(t)$, as
depicted in Fig.~\ref{fig:cs}.
 \setlength{\unitlength}{.09in}
\begin{figure}[h]
\begin{center}
\begin{picture}(78,31)(0,0)
%\put(0,0){\framebox(78,31){}}

 \put(5,0){
\put(5,10){{\ar{\filt{$s^*_p(-t)$}{\ar}}}} \put(21,10){
   \put(0,0){\line(3,2){3}}
   \qbezier(0,2)(2,2)(3,-1)
   \put(2.95,-1.3){\vector(0,-1){0.05}}}
\put(24,7){\makebox(0,0){$t=nT$}} \put(25,10){\ar}
\put(40,6.5){\ar{\rmult{\ar{\filt{$a_m(t)$}{\ar}}}}}
\put(27,11.5){\makebox(0,0){$y_p[n]$}}
\put(46,1.5){\makebox(0,0){$\sum_{n \in \ZZ} \delta(t-nT)$}}
\put(45,3){\vector(0,1){2.5}} \put(42,8){\makebox(0,0){$d_m[n]$}}
}

 \put(20,15){{\Large \vdots}} \put(50,15){{\Large \vdots}}

 \put(5,20){

 \put(5,2){{\ar{\filt{$s^*_1(-t)$}{\ar}}}} \put(21,2){
   \put(0,0){\line(3,2){3}}
   \qbezier(0,2)(2,2)(3,-1)
   \put(2.95,-1.3){\vector(0,-1){0.05}}}
\put(24,-1){\makebox(0,0){$t=nT$}} \put(25,2){\ar}
\put(40,6.5){\ar{\rmult{\ar{\filt{$a_1(t)$}{\ar}}}}}
\put(27,3.5){\makebox(0,0){$y_1[n]$}}
\put(46,1.5){\makebox(0,0){$\sum_{n \in \ZZ} \delta(t-nT)$}}
\put(45,3){\vector(0,1){2.5}} \put(42,8){\makebox(0,0){$d_1[n]$}}

}

\put(5,16){\st{$x(t)$}{\ar}} \put(10,10){\line(0,1){12}}
\put(34,4){\line(0,1){25}} \put(45,4){\line(0,1){25}}
\put(34,4){\line(1,0){11}} \put(34,29){\line(1,0){11}}
\put(40,17){\makebox(0,0){$\mbox{CTF}$}}
\put(66,16){\radder{\ar{\et{$x(t)$}}}}
\put(67,6.5){\vector(0,1){8.5}} \put(67,26.5){\vector(0,-1){9.3}}
\end{picture}
\end{center}
\caption{Compressed sensing of analog signals. The sampling
functions $s_i(t)$ are obtained by combining the blocks in
Fig.~\ref{fig:samp} and are given in Theorem~\ref{thm:fbcs}.}
\label{fig:cs}
\end{figure}
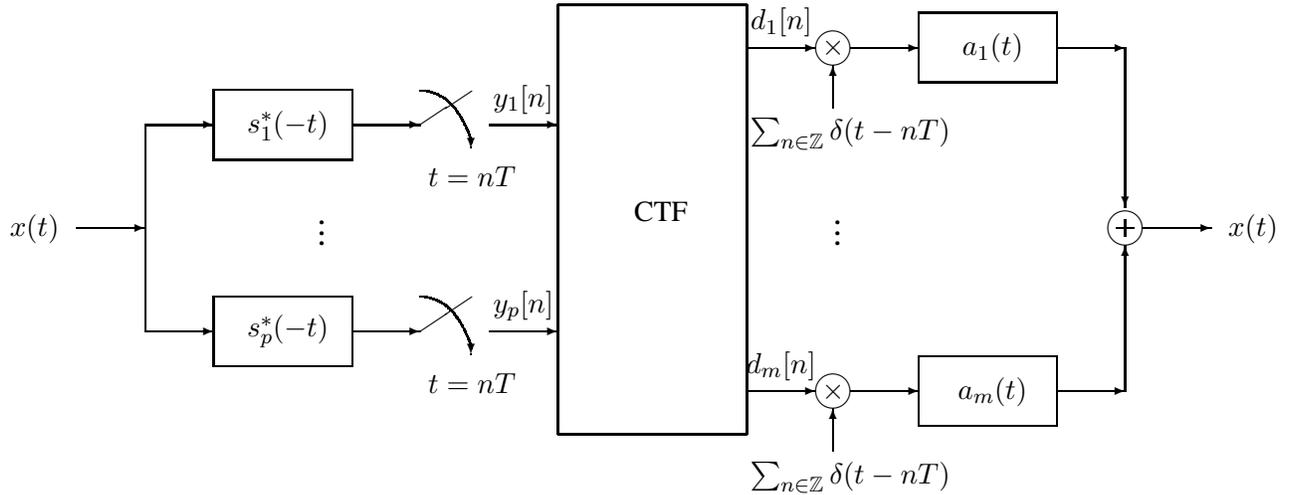

As a final comment, we note that we may add an invertible diagonal
matrix $\bbz\ejo$ prior to multiplication by $\bba$. Indeed, in
this case the measurements are given by
\begin{equation}
\label{eq:imvs2} \by(e^{j\omega })=\bbw\ejo\bba
\bbz(e^{j\omega})\bd(e^{j\omega})=\bba\tilde{\bd}(e^{j\omega}),
\end{equation}
where $\tilde{\bd}(e^{j\omega})$ has the same sparsity profile as
$\bd(e^{j\omega})$. Therefore, $\tilde{\bd}\ejo$ can be recovered
using the CTF block. In order to reconstruct $x(t)$, we first
filter each of the non-zero sequences $\tilde{d}_i[n]$ with the
convolutional inverse of $Z_i(e^{j\omega })$.

In this section we discussed the basic elements that allow
recovery from compressed analog signals: we first use a
biorthogonal sampling set in order to access the coefficient
sequences, and then employ a conventional CS mixing matrix to
compress the measurements. Recovery is possible by using an IMV
model and applying the CTF block of Fig.~\ref{FigIMV} either in
time or in frequency.  In practical applications we have the
freedom to choose $\bbw\ejo$ and $\bba$ so that we end up with
analog sampling functions that are easy to implement. Two examples
are considered in the next section.
\section{Examples} \label{sec:examples}

\subsection{Periodic Sparsity}

Suppose that we are given a signal $x(t)$ that lies in a SI
subspace generated by $a(t)$ so that $x(t)=\sum_{n \in \ZZ} d[n]
a(t-nT')$. The coefficients $d[n]$ have a periodic sparsity
pattern: Out of each consecutive group of $m$ coefficients, there
are at most $k$ non-zero values, in a given pattern. For example,
suppose that $m=7,k=2$ and the sparsity profile is $S=\{1,4\}$.
Then $d[n]$ can be non-zero only at indices $n=1+7\ell$ or
$n=4+7\ell$ for some integer $\ell$. Decomposing $d[n]$ into
blocks $\bd[n]$ of length $m$, the sparsity pattern of $x(t)$
implies that the vectors $\{\bd[n]\}$ are jointly $k$-sparse.

Since $x(t)$ lies in a SI subspace spanned by a single generator,
we can sample it by first prefiltering with the filter
\begin{equation}
\label{eq:sex} Q(\omega)=\frac{1}{\phi_{HA}(e^{j\omega
T'})}H(\omega),
\end{equation}
where $h(t)$ is any function such that $\phi_{HA}\ejo$ defined by
(\ref{eq:psa}) is non-zero a.e. on $\omega$, and then sampling the
output at rate $1/T'$, as in Fig.~\ref{fig:sampling}. With this
choice, the samples $c[n]=\inner{q(t-nT')}{x(t)}$ are equal to the
unknown coefficients $d[n]$. We may then use standard CS
techniques to compressively sample $d[n]$. For example, we can
sample $d[n]$ sequentially by considering blocks $\bd[n]$ of
length $m$, and using a standard CS matrix $\bba$ designed to
sample a $k$-sparse vector of length-$m$. Alternatively, we can
exploit the joint sparsity by combining several blocks and
sampling them together using MMV techniques, or applying the IMV
method of Section~\ref{sec:imv}. However, these approaches still
require an analog sampling rate of $1/T'$. Thus, the rate
reduction is only in discrete time, whereas the analog sampling
rate remains at the Nyquist rate. Since many of the coefficients
are zero, we would like to directly reduce the analog rate so as
not to acquire the zero values of $d[n]$, rather than acquiring
them first, and then compressing in discrete time.

To this end, we note that our problem may be viewed as a special
case of the general model (\ref{eq:model}) with
$a_i(t)=a(t-(i-1)),1 \leq i \leq m$ and $d_i[n]=d[i+nT],1 \leq i
\leq m$ with $T=mT'$. Therefore, the rate can be reduced by using
the tools of Section~\ref{sec:acs}. From Theorem~\ref{thm:fbcs},
we first need to construct a set of functions $\{v_\ell(t-nT)\}$
that are biorthogonal to $\{a_\ell(t-nT)\}$. It is easy to see
that
\begin{equation}
v_{\ell}(t)=q(t-(\ell-1)),\quad 1 \leq \ell \leq m,
\end{equation}
with $Q(\omega)$ given by (\ref{eq:sex}) constitute a biorthogonal
set. Indeed, with this choice
\begin{eqnarray}
\label{eq:has}\lefteqn{\hspace*{-0.2in}
[\bbm_{VA}\ejo]_{i\ell}=\frac{1}{T} e^{j(i-\ell)\omega/T} \cdot }
\nonumber \\
&& \hspace*{-0.35in} \sum_{k \in \ZZ} e^{-j(i-\ell) 2\pi k/T} Q^*
\bl \frac{\omega}{T} -\frac{2\pi }{T}k \br A \bl \frac{\omega}{T}
-\frac{2\pi }{T}k \br \nonumber \\
&& \hspace*{-0.35in} =  \frac{1}{T}e^{j(i-\ell)\omega/T}
\sum_{r=0}^{T-1} e^{-j(i-\ell) 2\pi r/T} G(e^{j(\omega-2\pi r)/T}
),
\end{eqnarray}
where we defined
\begin{equation}
\label{eq:g} G(e^{j\omega} )=\frac{1}{T}\sum_{k \in \ZZ} Q^* \bl
\omega -2\pi
 k \br A \bl \omega -2\pi k \br.
\end{equation}
From (\ref{eq:sex}), $G(e^{j\omega} )=1$. Combining this with the
relation
\begin{equation}
\frac{1}{T}\sum_{r=0}^{T-1} e^{-j2\pi r m/T}=\delta[m],
\end{equation}
it follows from (\ref{eq:has}) that $\bbm_{VA}\ejo=\bbi$.

We now use Theorem~\ref{thm:fbcs} to conclude that any sampling
functions of the form $\bs(\omega)=\bbw^*(e^{j\omega
T})\bba^*\bv(\omega)$ with
$V_\ell(\omega)=Q(\omega)e^{-j(\ell-1)\omega}$ and $Q(\omega)$
given by (\ref{eq:sex}) can be used to compressively sample the
signal at a rate $p/T=p/(mT')<1/T'$. In particular, given a matrix
$\bba$ of size $p \times m$ that satisfies the CS requirements, we
may choose sampling functions
\begin{equation}
\label{eq:sex1} s_i(t)=\sum_{\ell=1}^m \bba^*_{i\ell}
q(t-(\ell-1)),\quad 1 \leq i \leq p.
\end{equation}
In this strategy, each sample is equal to a linear combination of
several values $d[n]$, in contrast to the high-rate method in
which each sample is exactly equal $d[n]$.

As a special case, suppose that $T'=1$, $a(t)=1$ on the interval
$[0,1]$ and is zero otherwise. Thus, $x(t)$ is piecewise constant
over intervals of length $1$. Choosing $h(t)=a(t)$ in
(\ref{eq:sex}) it follows that $q(t)=1$. This is because
$r_{aa}[n]=\inner{a(t-n)}{a(t)}=\delta[n]$ so that
$\phi_{AA}\ejo=1$. Therefore, $v_\ell(t)=a(t-(\ell-1))$ are
biorthogonal to $a_{\ell}(t)$. One way to acquire the coefficients
$d[n]$ is to filter the signal $x(t)$ with $v(t)=a(t)$ and sample
the output at rate $1$. This corresponds to integrating $x(t)$
over intervals of length one. Since $x(t)$ has a constant value
$d[n]$ over the $n$th interval, the output will indeed be the
sequence $d[n]$. To reduce the rate, we may instead use the $p<m$
sampling functions (\ref{eq:sex1}) and sample the output at rate
$1/m$.  This is equivalent to first multiplying $x(t)$ by $p$
periodic sequences with period $m$. Each sequence is piecewise
constant over intervals of length $1$ with values $\bba_{i\ell},1
\leq \ell \leq m$. The continuous output is then integrated over
intervals of length $m$ to produce the samples $y_{\ell}[n]$.
Applying the CTF block to these measurements allows to recover
$d[n]$. Although this special case is rather simple, it highlights
the main idea. Furthermore, the same techniques can be used even
when the  generator $a(t)$ has infinite length.

\subsection{Multiband Sampling} Consider next the multiband
sampling problem \cite{ME07,ME09} in which we have a complex
signal that consists of at most $N$ frequency bands, each of
length no larger than $B$. In addition, the signal is bandlimited
to $2\pi/T$. If the band locations are known, then we can recover
such a signal from nonuniform samples at an average rate of
$NB/(2\pi)$ which is typically much smaller than the Nyquist rate
$2\pi/T$ \cite{LV98,HW99,VB00}. When the band locations are
unknown, the problem is much more complicated. In \cite{ME07} it
was shown that the minimum sampling rate for such signals is
$NB/\pi$. Furthermore, explicit algorithms were developed which
achieve this rate. Here we illustrate how this problem can be
formulated within our framework.

Dividing the frequency interval $[0,2\pi/T)$ into $m$ sections,
each of equal length $2\pi/(mT)$, it follows that if $m \leq
2\pi/(BT)$ then each band comprising the signal is contained in no
more than $2$ intervals. Since there are $N$ bands, this implies
that at most $2N$ sections contain energy. To fit this problem
into our general model, let
\begin{equation}
A_i(\omega)=\sqrt{mT} A\bl \omega -\frac{2\pi (i-1)}{mT} \br,\quad
1 \leq i \leq m,
\end{equation}
where $A(\omega)$ is a low-pass filter (LPF) on $[0,2\pi/(mT))$.
Thus, $A_i(\omega)$ describes the support of the $i$th interval.
Since any multiband signal $x(t)$ is supported in the frequency
domain over at most $2N$ sections, $x(t)$ can be written as
\begin{equation}
X(\omega)=\sum_{i=1}^{m} A_i(\omega) D_i(\omega),
\end{equation}
for some $D_i(\omega)$ supported on the $i$th interval, where at
most $2N$ functions are nonzero. Since the support of
$D_i(\omega)$ has length $2\pi/(mT)$, it can be written as a
Fourier series
\begin{equation}
D_i(\omega)=\sum_{n \in \ZZ} d_i[n]e^{-j\omega nmT} \deft D_i(e^{j
\omega mT}).
\end{equation}
Thus, our signal fits the general model (\ref{eq:model}), where
there are at most $2N$ sequences $d_i[n]$ that are nonzero.

We now use our general results to obtain sampling functions that
can be used to sample and recover such signals at rates lower than
Nyquist. One possibility is to choose $h_i(t)=a_i(t)$. Since the
functions $a_i(t)$ are orthonormal (as is evident by considering
the frequency domain representation), we have that
$\bbm_{HA}\ejo=\bbm_{AA}\ejo=\bbi$. Consequently, the resulting
sampling functions are
\begin{equation}
s_i(t)=\sum_{\ell=1}^m \bba_{i\ell}^*a_{\ell}(t).
\end{equation}
In the Fourier domain, $S_i(\omega)$ is bandlimited to $2\pi/T$
and piecewise constant with values $\sqrt{mT}\bba_{i\ell}^*$ over
intervals of length $2\pi/(mT)$.

Alternative sampling functions are those used in \cite{ME07}:
\begin{equation}
\label{eq:delay} s_i(t)=\delta(t-c_iT),\quad 1 \leq i \leq p,
\end{equation}
where $\{c_i\}$ are $p$ distinct integer values in the range $1
\leq c_i \leq m$. Since $x(t)$ is bandlimited, sampling with the
filters (\ref{eq:delay}) is equivalent to using the bandlimited
functions
 \begin{equation}
\label{eq:dbl} S_i(\omega)= e^{-j c_i \omega T},\quad 0 \leq
\omega \leq \frac{2\pi}{T}.
 \end{equation}
To show that these filters can be obtained from our general
framework incorporated in Theorem~\ref{thm:fbcs}, we need to
choose a $p \times m$ matrix $\bba$ and an invertible $p \times p$
matrix $\bbw\ejo$ such that $G_i(\omega)=S_i(\omega)$ where
\begin{equation}
\label{eq:go} \bg(\omega)=\bbw^*(e^{j\omega mT})\bba^*\bv(\omega),
\end{equation}
and $\bv(\omega)$ represents a biorthogonal set. In our setting,
we can choose $V_i(\omega)=A_i(\omega)$ due to the orthogonality
of $a_i(t)$.

Let $\bba$ be the matrix consisting of the rows $c_i,1 \leq i \leq
p$ of the $m \times m$ Fourier matrix
\begin{equation}
\bba_{i\ell}=\frac{1}{\sqrt{m}}e^{j2\pi (\ell-1)c_i/m},
\end{equation}
and choose $\bbw\ejo$ as a diagonal matrix with $i$th diagonal
element
\begin{equation}
W_i\ejo=\frac{1}{\sqrt{T}}e^{j c_i\omega/m},\quad 0 \leq \omega <
2\pi.
\end{equation}
From (\ref{eq:go}),
\begin{equation}
G_i(\omega)=W_i^*(e^{j\omega mT})\sum_{\ell=1}^m
\bba_{i\ell}^*A_\ell(\omega).
\end{equation}
Since $A_\ell(\omega)$ is equal to $\sqrt{mT}$ over the $\ell$th
interval $[(\ell-1)2\pi/(mT),\ell 2\pi/mT)$ and $0$ otherwise,
$\sum_{\ell=1}^m \bba_{i\ell}^*A_\ell(\omega)$ is piecewise
constant with values equal to $\sqrt{mT}\bba_{i\ell}^*$ on
intervals of length $2\pi/(mT)$. In addition, on the $\ell$th
interval,
\begin{eqnarray}
W_i^*(e^{j\omega mT})& = & \frac{1}{\sqrt{T}}e^{-j
c_iT(\omega-(\ell-1)2\pi/(mT)) } \nonumber \\
& = & (\sqrt{m}/\sqrt{T})e^{-j c_i \omega T}\bba_{i\ell}.
\end{eqnarray}
Consequently, on this interval, $G_i(\omega)$  is equal to
$\sqrt{mT}W_i^*(e^{j\omega mT})\bba^*_{i\ell}= e^{-jc_i \omega
T}$. Since this expression does not depend on $\ell$,
$G_i(\omega)=S_i(\omega)$ for $0 \leq \omega \leq 2\pi/T$.

From our general results, in order to recover the original  signal
$x(t)$ we need to apply the CTF to the modified measurements
$\tilde{\by}\ejo=\bbw^{-1}\ejo\by\ejo$. Since $\bbw\ejo$ is
diagonal, the DTFT of the $i$th sequence $\tilde{y}_i[n]$ is given
by
\begin{equation}
\tilde{Y}_i\ejo=\frac{1}{\sqrt{T}}e^{-jc_i\omega/m}Y_i\ejo,\quad 0
\leq \omega \leq 2\pi.
\end{equation}
This corresponds to a scaled non-integer delay $c_i/m$ of
$y_i[n]$. Such a delay can be realized by first upsampling the
sequence $y_i[n]$ by factor of $m$, low-pass filtering with a LPF
with cut-off $\pi/m$, shifting the resulting sequence by $c_i$,
and then down sampling by $m$. This coincides with the approach
suggested in \cite{ME07} for applying the CTF directly in the time
domain. Here we see that this processing follows directly from our
general framework.

We have shown that a particular choice of $\bba$ and $\bbw\ejo$
results in the sampling strategy of \cite{ME07}. Alternative
selections can lead to a variety of different sampling functions
for the same problem. The added value in this context is that in
\cite{ME07} there is no discussion on what type of sampling
methods lead to stable recovery. The framework we developed in
this paper can be applied in this specific setting to suggest more
general types of stable sampling and recovery strategies.

\section{Conclusion}

We developed a general framework to treat sampling of sparse
analog signals. We focused on signals in a SI space generated by
$m$ kernels, where only $k$ out of the $m$ generators are active.
The difficulty arises from the fact that we do not know in advance
which $k$ are chosen. Our approach was based on merging ideas from
standard analog sampling, with results from the emerging field of
CS. The latter focuses on sensing finite-dimensional vectors that
have a sparsity structure in some transform domain. Although our
problem is inherently infinite-dimensional, we showed that by
using the notion of biorthogonal sampling sets and the recently
developed CTF block \cite{ME08,ME07}, we can convert our problem
to a finite-dimensional counterpart that takes on the form of an
MMV, a problem which has been treated previously in the CS
literature.

In this paper, we focused on sampling using a bank of analog
filters. An interesting future direction to pursue is to extend
these ideas to other sampling architectures that may be easier to
implement in hardware.

As a final note, most of the literature to date on the exciting
field of CS has focused on sensing of finite-dimensional vectors.
On the other hand, traditional sampling theory focuses on
infinite-dimensional continuous-time signals. It is our hope that
this work can serve as a step in the direction of merging these
two important areas in sampling, leading to a more general notion
of compressive sampling.

\section{Acknowledgement}

The author would like to thank Moshe Mishali for many fruitful
discussions, and the reviewers for useful comments on the
manuscript which helped improve the exposition.

%\bibliographystyle{IEEEbib}
%\bibliography{notes,paper,moshiko_bib}

\begin{thebibliography}{10}

\bibitem{S49}
C.~E. Shannon,
\newblock ``Communications in the presence of noise,''
\newblock {\em Proc. IRE}, vol. 37, pp. 10--21, Jan 1949.

\bibitem{N28}
H.~Nyquist,
\newblock ``Certain topics in telegraph transmission theory,''
\newblock {\em EE Trans.}, vol. 47, pp. 617--644, Jan. 1928.

\bibitem{D90}
I.~Daubechies,
\newblock ``The wavelet transform, time-frequency localization and signal
  analysis,''
\newblock {\em IEEE Trans. Inform. Theory}, vol. 36, pp. 961--1005, Sep. 1990.

\bibitem{U00}
M.~Unser,
\newblock ``Sampling---50 years after {S}hannon,''
\newblock {\em IEEE Proc.}, vol. 88, pp. 569--587, Apr. 2000.

\bibitem{AU94}
A.~Aldroubi and M.~Unser,
\newblock ``Sampling procedures in function spaces and asymptotic equivalence
  with {S}hannon's sampling theory,''
\newblock {\em Numer. Funct. Anal. Optimiz.}, vol. 15, pp. 1--21, Feb. 1994.

\bibitem{UA94}
M.~Unser and A.~Aldroubi,
\newblock ``A general sampling theory for nonideal acquisition devices,''
\newblock {\em IEEE Trans. Signal Processing}, vol. 42, no. 11, pp. 2915--2925,
  Nov. 1994.

\bibitem{V01}
P.~P. Vaidyanathan,
\newblock ``Generalizations of the sampling theorem: Seven decades after
  {N}yquist,''
\newblock {\em IEEE Trans. Circuit Syst. I}, vol. 48, no. 9, pp. 1094--1109,
  Sep. 2001.

\bibitem{E02}
Y.~C. Eldar,
\newblock ``Sampling and reconstruction in arbitrary spaces and oblique dual
  frame vectors,''
\newblock {\em J. Fourier Analys. Appl.}, vol. 1, no. 9, pp. 77--96, Jan. 2003.

\bibitem{ED04}
Y.~C. Eldar and T.~G. Dvorkind,
\newblock ``A minimum squared-error framework for generalized sampling,''
\newblock {\em IEEE Trans. Signal Processing}, vol. 54, no. 6, pp. 2155--2167,
  Jun. 2006.

\bibitem{DEM08}
T.~G. Dvorkind, Y.~C. Eldar, and E.~Matusiak,
\newblock ``Nonlinear and non-ideal sampling: Theory and methods,''
\newblock {\em IEEE Trans. Signal Processing}, vol. 56, no. 12, pp. 5874--5890,
  Dec. 2008.

\bibitem{EM08}
Y.~C. Eldar and T.~Michaeli,
\newblock ``Beyond bandlimited sampling: Nonlinearities, smoothness and
  sparsity,''
\newblock to appear in {\em IEEE Signal Proc. Magazine}.

\bibitem{DDR94}
C.~de~Boor, R.~DeVore, and A.~Ron,
\newblock ``{The structure of finitely generated shift-invariant spaces in
  $L_2(\mathbb{R}^d)$},''
\newblock {\em J. Funct. Anal}, vol. 119, no. 1, pp. 37--78, 1994.

\bibitem{GHM94}
J.~S. Geronimo, D.~P. Hardin, and P.~R. Massopust,
\newblock ``{Fractal functions and wavelet expansions based on several scaling
  functions},''
\newblock {\em Journal of Approximation Theory}, vol. 78, no. 3, pp. 373--401,
  1994.

\bibitem{CE04}
O.~Christansen and Y.~C. Eldar,
\newblock ``Oblique dual frames and shift-invariant spaces,''
\newblock {\em Appl. Comp. Harm. Anal.}, vol. 17, no. 1, pp. 48--68, 2004.

\bibitem{CE05}
O.~Christensen and Y.~C. Eldar,
\newblock ``Generalized shift-invariant systems and frames for subspaces,''
\newblock {\em J. Fourier Analys. Appl.}, vol. 11, pp. 299--313, 2005.

\bibitem{AG01}
A.~Aldroubi and K.~Gr{\"o}chenig,
\newblock ``Non-uniform sampling and reconstruction in shift-invariant
  spaces,''
\newblock {\em {\em Siam Review}}, vol. 43, pp. 585--620, 2001.

\bibitem{S73b}
I.~J. Schoenberg,
\newblock {\em Cardinal Spline Interpolation},
\newblock Philadelphia, PA: SIAM, 1973.

\bibitem{LV98}
Y.-P. Lin and P.~P. Vaidyanathan,
\newblock ``Periodically nonuniform sampling of bandpass signals,''
\newblock {\em IEEE Trans. Circuits Syst. II}, vol. 45, no. 3, pp. 340--351,
  Mar. 1998.

\bibitem{HW99}
C.~Herley and P.~W. Wong,
\newblock ``Minimum rate sampling and reconstruction of signals with arbitrary
  frequency support,''
\newblock {\em IEEE Trans. Inform. Theory}, vol. 45, no. 5, pp. 1555--1564,
  July 1999.

\bibitem{VB00}
R.~Venkataramani and Y.~Bresler,
\newblock ``Perfect reconstruction formulas and bounds on aliasing error in
  sub-nyquist nonuniform sampling of multiband signals,''
\newblock {\em IEEE Trans. Inform. Theory}, vol. 46, no. 6, pp. 2173--2183,
  Sep. 2000.

\bibitem{ME07}
M.~Mishali and Y.~C. Eldar,
\newblock ``Blind multi-band signal reconstruction: Compressed sensing for
  analog signals,''
\newblock {\em IEEE Trans. Signal Process.}, vol. 57, pp. 993--1009, Mar. 2009.

\bibitem{ME08c}
M.~Mishali and Y.~C. Eldar,
\newblock ``Spectrum-blind reconstruction of multi-band signals,''
\newblock in {\em Proc. Int. Conf. Acoust., Speech, Signal Processing
  (ICASSP-2008), (Las Vegas, USA)}, April 2008, pp. 3365--3368.

\bibitem{ME09}
M.~Mishali and Y.~C. Eldar,
\newblock ``From theory to practice: Sub-{N}yquist sampling of sparse wideband
  analog signals,''
\newblock {\em arXiv 0902.4291; {\em submitted to } IEEE Selcted Topics on
  Signal Process.}, 2009.

\bibitem{LD08}
Y.~M. Lu and M.~N. Do,
\newblock ``A theory for sampling signals from a union of subspaces,''
\newblock {\em IEEE Trans. Signal Processing}, vol. 56, no. 6, pp. 2334--2345,
  2008.

\bibitem{BD09}
T.~Blumensath and M.~E. Davies,
\newblock ``Sampling theorems for signals from the union of finite-dimensional
  linear subspaces,''
\newblock {\em IEEE Trans. Inform. Theory},
\newblock to appear.

\bibitem{EM082}
Y.~C. Eldar and M.~Mishali,
\newblock ``Robust recovery of signals from a union of subspaces,''
\newblock {\em {\em submitted to } IEEE Trans. Inform. Theory}, July 2008.

\bibitem{D06}
D.~L. Donoho,
\newblock ``Compressed sensing,''
\newblock {\em {IEEE} Trans. on Inf. Theory}, vol. 52, no. 4, pp. 1289--1306,
  Apr 2006.

\bibitem{CRT06}
E.~J. Cand\`{e}s, J.~Romberg, and T.~Tao,
\newblock ``Robust uncertainty principles: Exact signal reconstruction from
  highly incomplete frequency information,''
\newblock {\em IEEE Trans. Inform. Theory}, vol. 52, no. 2, pp. 489--509, Feb.
  2006.

\bibitem{MPref}
S.~G. Mallat and Z.~Zhang,
\newblock ``Matching pursuits with time-frequency dictionaries,''
\newblock {\em IEEE Trans. Signal Processing}, vol. 41, no. 12, pp. 3397--3415,
  Dec. 1993.

\bibitem{BPref2}
S.~S. Chen, D.~L. Donoho, and M.~A. Saunders,
\newblock ``Atomic decomposition by basis pursuit,''
\newblock {\em SIAM J. Scientific Computing}, vol. 20, no. 1, pp. 33--–61,
  1999.

\bibitem{MEGref1}
I.~F. Gorodnitsky, J.~S. George, and B.~D. Rao,
\newblock ``{Neuromagnetic source imaging with FOCUSS: A recursive weighted
  minimum norm algorithm},''
\newblock {\em J. Electroencephalog. Clinical Neurophysiol.}, vol. 95, no. 4,
  pp. 231–--251, Oct. 1995.

\bibitem{MElad}
D.~L. Donoho and M.~Elad,
\newblock ``Maximal sparsity representation via $\ell 1$ minimization,''
\newblock {\em Proc. Natl. Acad. Sci.}, vol. 100, pp. 2197--–2202, Mar. 2003.

\bibitem{CT05}
E.~J. Cand\`{e}s and T.~Tao,
\newblock ``Decoding by linear programming,''
\newblock {\em IEEE Trans. Inform. Theory}, vol. 51, no. 12, pp. 4203--4215,
  Dec. 2005.

\bibitem{Analog2Info1}
J.~A. Tropp, M.~B. Wakin, M.~F. Duarte, D.~Baron, and R.~G.
Baraniuk,
\newblock ``Random filters for compressive sampling and reconstruction,''
\newblock in {\em Proc. IEEE International Conference on Acoustics, Speech and
  Signal Processing ICASSP 2006}, May 2006, vol.~3.

\bibitem{Analog2Info2}
J.~N. Laska, S.~Kirolos, M.~F. Duarte, T.~S. Ragheb, R.~G.
Baraniuk, and
  Y.~Massoud,
\newblock ``Theory and implementation of an analog-to-information converter
  using random demodulation,''
\newblock in {\em Proc. IEEE International Symposium on Circuits and Systems
  ISCAS 2007}, May 2007, pp. 1959--1962.

\bibitem{VMB02}
M.~Vetterli, P.~Marziliano, and T.~Blu,
\newblock ``Sampling signals with finite rate of innovation,''
\newblock {\em {IEEE} Trans. Signal Processing}, vol. 50, pp. 1417--1428, June
  2002.

\bibitem{DVB07}
P.L. Dragotti, M.~Vetterli, and T.~Blu,
\newblock ``Sampling moments and reconstructing signals of finite rate of
  innovation: {S}hannon meets {S}trang-{F}ix,''
\newblock {\em {IEEE} Trans. Sig. Proc.}, vol. 55, no. 5, pp. 1741--1757, May
  2007.

\bibitem{ME08}
M.~Mishali and Y.~C. Eldar,
\newblock ``Reduce and boost: Recovering arbitrary sets of jointly sparse
  vectors,''
\newblock {\em IEEE Trans. Signal Process.}, vol. 56, no. 10, pp. 4692--4702,
  Oct. 2008.

\bibitem{EC04}
Y.~C. Eldar and O.~Christansen,
\newblock ``Characterization of oblique dual frame pairs,''
\newblock {\em J. Applied Signal Processing}, pp. 1--11, 2006,
\newblock Article ID 92674.

\bibitem{Chen}
J.~Chen and X.~Huo,
\newblock ``Theoretical results on sparse representations of
  multiple-measurement vectors,''
\newblock {\em IEEE Trans. Signal Processing}, vol. 54, no. 12, pp. 4634--4643,
  Dec. 2006.

\bibitem{Kruskal}
J.~B. Kruskal,
\newblock ``Three-way arrays: Rank and uniqueness of trilinear decompositions,
  with application to arithmetic complexity and statistics,''
\newblock {\em Linear Alg. Its Applic.}, vol. 18, no. 2, pp. 95--138, 1977.

\bibitem{C08}
E.~J. Cand\`{e}s,
\newblock ``The restricted isometry property and its implications for
  compressed sensing,''
\newblock {\em C. R. Acad. Sci. Paris, Ser. I}, vol. 346, pp. 589--592, 2008.

\bibitem{CR07}
E.~J. Cand\`{e}s and J.~Romberg,
\newblock ``Sparsity and incoherence in compressive sampling,''
\newblock {\em Inverse Prob.}, vol. 23, no. 3, pp. 969--985, 2007.

\bibitem{Cotter}
S.~F. Cotter, B.~D. Rao, K.~Engan, and K.~Kreutz-Delgado,
\newblock ``Sparse solutions to linear inverse problems with multiple
  measurement vectors,''
\newblock {\em IEEE Trans. Signal Processing}, vol. 53, no. 7, pp. 2477--2488,
  July 2005.

\end{thebibliography}

\end{document}